\documentclass[]{iopart}
\usepackage{graphicx,subfigure,bm,txfonts}
\usepackage{iopams}
\usepackage{setstack}
\usepackage{overpic}
\usepackage{pifont}

\newcommand{\kp}{k_\varphi}
\newcommand{\kz}{k_z}
\newcommand{\kv}{{\bf k}}
\newcommand{\qv}{{\bf q}}

\makeatletter
\newcommand{\@textsubscript}[1]{%
{\m@th\ensuremath{_{\mbox{\fontsize\sf@size\z@#1}}}}}
\DeclareRobustCommand*\textsubscript[1]{%
\@textsubscript{\selectfont#1}}
\makeatother

\begin{document}

\title{Flux Periodicities in Loops of Nodal Superconductors}

\author{Florian Loder}
\author{Arno P. Kampf}
\author{Thilo Kopp}
\author{Jochen Mannhart\\[0.6cm]}
\address{
Center for Electronic Correlations and Magnetism, Institute of Physics, University of 
Augsburg, D-86135 Augsburg, Germany}

\date{\today}

\begin{abstract}
Supercurrents in superconducting flux threaded loops are expected to oscillate with the magnetic flux with a period of $hc/2e$. This is indeed true for $s$-wave superconductors larger than the coherence length $\xi_0$. Here we show that for superconductors with gap nodes, there is no such strict condition for the supercurrent to be $hc/2e$ rather than $hc/e$ periodic. For nodal superconductors, the flux induced Doppler shift of the near nodal states leads to a flux dependent occupation probability of quasi-particles circulating clockwise and counter clockwise around the loop, which leads to an $hc/e$ periodic component of the supercurrent, even at zero temperature. We analyze this phenomenon on a cylinder in an approximative analytic approach and also numerically within the framework of the BCS theory. Specifically for $d$-wave pairing, we show that the $hc/e$ periodic current component decreases with the inverse radius of the loop and investigate its temperature dependence.
\end{abstract}

\pacs{74.20.Fg, 74.25.Fy, 74.25.Sv}

\maketitle

\section{Introduction} \label{sec1}

Electrons moving in a multiply connected geometry threaded by a magnetic flux $\Phi$ are an ideal system to observe quantum mechanical phase coherence. If an electron encircles a flux threaded hole on a closed path, the phase difference of its wave function must be a multiple of $2\pi$ plus the Aharonov-Bohm phase $2\pi\,\Phi/\Phi_0$ where $\Phi_0=hc/e$ is the flux quantum \cite{AB}. Therefore, a finite phase gradient persists in the wave function for all flux values $\Phi/\Phi_0\neq1$. Consequently, a persistent current is flowing around the hole, which is modulated by the magnetic flux with a period of $\Phi_0$ \cite{Landauer, landauer:85}.

This phenomenon is best observed in phase coherent superconducting (SC) rings \cite{London,Byers,schrieffer}. Measurements of magnetic flux trapped in SC rings showed that the flux is quantized in multiples of $\Phi_0/2$ \cite{Doll,Deaver}, which implies a flux periodicity of $hc/2e$ for the circulating supercurrent and likewise for all thermodynamic quantities \cite{degennes}. Indeed, the same periodicity has been found by Little and
Parks in measurements of the critical temperature $T_c$ of flux threaded cylinders \cite{Little,Parks}. 

The $hc/2e$ flux periodicity of the SC state is naturally contained in the BCS pairing theory of superconductivity \cite{bcs}, as was shown by Byers and Yang \cite{Byers} and independently by Brenig \cite{brenig:61} and by Onsager \cite{onsager:61}. Byers and Yang introduced two distinct classes of SC wave functions, which are not related by a gauge transformation; one class of states has minima in the free energy at even multiples of the SC flux quantum $\Phi_0/2$, whereas the second class has minima at odd multiples of $\Phi_0/2$. They proved that the minima in the free energy become degenerate in the thermodynamic limit and all thermodynamic quantities $hc/2e$ periodic. In finite systems, this degeneracy is lifted. Consequences for flux-dependent oscillations of $T_c$ have been investigated by Bogachek \etal \cite{bogachek:75} using a quasi one dimensional (1D) thin-ring model with $s$-wave pairing. This and other recent works, including an analysis of the supercurrent, made evident that $s$-wave rings smaller than the SC coherence length $\xi_0$ display a flux periodicity of $hc/e$ rather than the anticipated $hc/2e$ periodicity \cite{czajka:05,wei:07,vakaryuk:08,loder:08.2}. Special attention arose from numerical investigations of SC $d$-wave loops \cite{loder:08,zhu:08}, in which the distinction of the two classes of SC states is more pronounced than in $s$-wave rings. These results agree with the analytical approaches using the thin-ring model \cite{barash:07,juricic:07} and will be extended in this article towards a two dimensional (2D) multi-channel model.

\begin{figure}[tb]
\centering
\includegraphics[width=8.4cm]{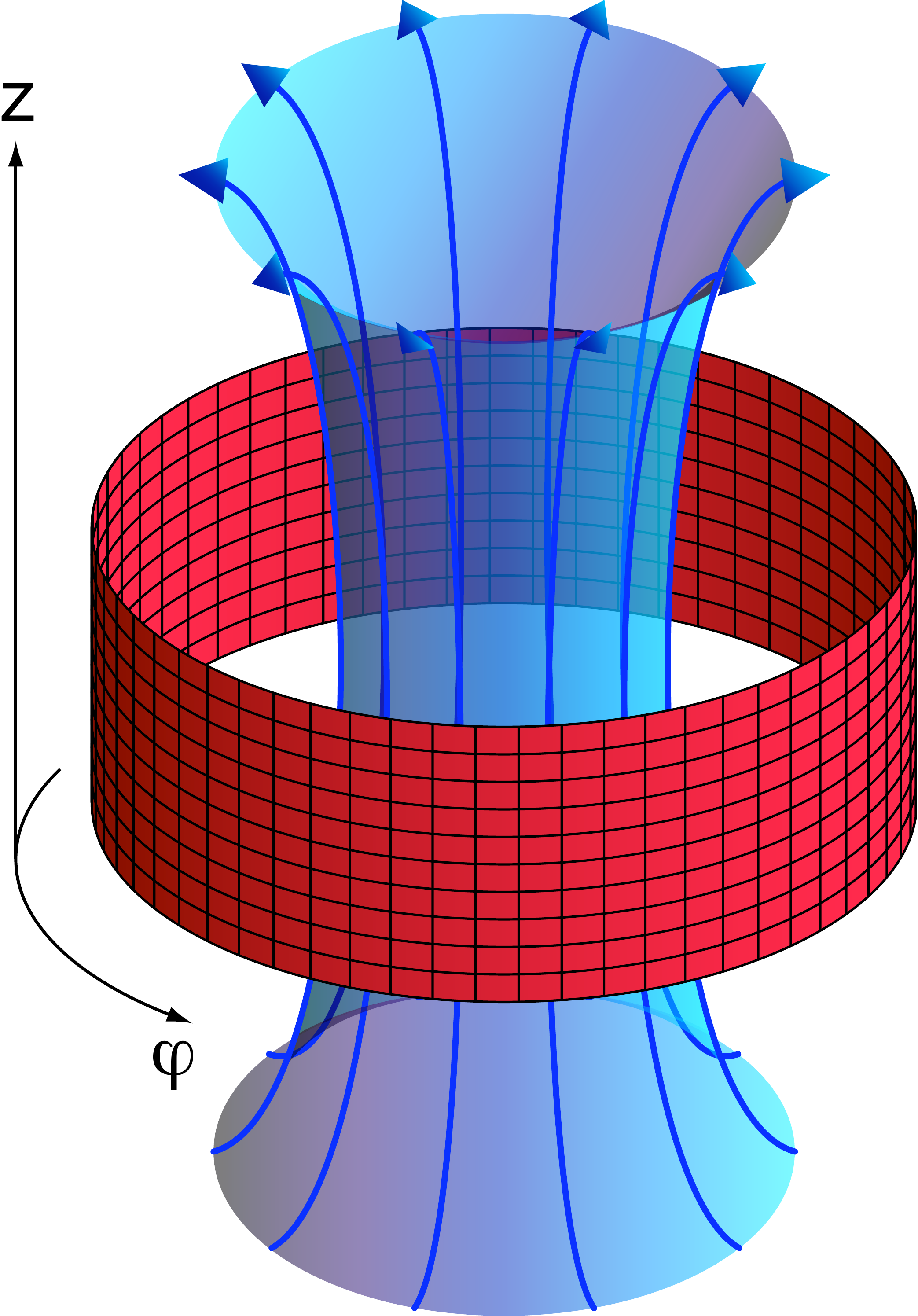}
\caption{As a model system to study persistent supercurrents we use a thin-wall cylinder constructed of a $2D$ discrete lattice. The interior of the cylinder is threaded by a magnetic flux $\Phi$; we assume that the flux does not penetrate into the cylinder itself. In such a system, $\Phi$ can be chosen arbitrarily, since quantization applies to the fluxoid and not the flux itself.}
\label{Fig0}
\end{figure}

For the discussion of the flux periodicity of the supercurrent we choose a discrete 2D lattice in a cylindrical geometry (figure~\ref{Fig0}). We recall that the magnetic flux threading a superconducting loop is quantized in units of the superconducting flux quantum $hc/2e$ \cite{Doll, Deaver}. This quantization reflects the minima of the free energy \cite{Byers}. These minima are determined by gauge invariance and the electron interaction; the flux quantum $hc/2e$ is therefore a fundamental property of any superconductor. Flux quantization in a cylinder requires that its walls are thicker than the penetration depth $\lambda$. If the walls are thinner than $\lambda$, the cylinder can be threaded by an arbitrary magnetic flux and only the quantity called fluxoid is quantized \cite{London, degennes, schrieffer}. In this situation, it is the flux periodicity of thermodynamical quantities such as the supercurrent or $T_c$, for which the pairing of electrons suggests $hc/2e$ periodicity.

For two reasons we expect nodal rather than nodeless superconductors to support an $hc/e$ periodicity. The first arises from the discrete nature of the eigenenergies in a finite system. The results of the summation over occupied eigenstates for integer and half-integer flux values differ by an amount proportional to the mean level spacing $\delta_F$ in the vicinity of the Fermi energy $E_F$. In the normal state, $\delta_F\propto1/V$, where $V$ is the volume of the system. For the thin cylinder shown in figure~\ref{Fig0} with a circumference $Na$ and a height $Ma$, where $a$ is the lattice constant, the level spacing is $\delta_F\propto1/(NM)$; in $s$-wave superconductors with an order parameter $\Delta\gg\delta_F$, $\delta_F$ matters little. For SC states with gap nodes, the situation is different. For example, in the $d$-wave superconductors with an order parameter $\Delta_\kv\propto \kp^2-\kz^2$, the nodal states closest to $E_F$ have to fulfill the condition $\kz=\kp$, thus there are fewer possible eigenstates and $\delta_F\propto1/N$.

The second reason is that for gapless superconductors with a finite density of states (DOS) close to $E_F$, the occupation probabilities of these states change with flux. The flux dependence of the occupation enhances the difference of current matrix elements for integer and half-integer flux values \cite{loder:08,loder:08.2,juricic:07}. This effect can be understood in terms of the spacial extension of a Cooper pair. In $s$-wave superconductors, the occupation probability remains constant for all $\Phi$, if the diameter of the cylinder is larger than $\xi_0$. If this condition is fulfilled, the constituents of a Cooper pair cannot circulate separately; the pair does not \textquotedblleft feel\textquotedblright\ the multiply connected geometry of the cylinder. But for nodal SC states, the length scale which characterizes their coherence, diverges in the nodal directions and there are always Cooper pairs which extend around the circumference of the cylinder. Therefore nodal superconductors have no characteristic length scale above which the SC state is unaffected by the geometry of the system.
These two combined effects are investigated on the basis of an analytical model in section~\ref{sec3} and by numerical calculations in section~\ref{sec4}.

\section{Superconductivity in a Flux-Threaded Cylinder} \label{sec2}
\begin{figure}[tb]
\centering
\begin{overpic}
[width=8.4cm]{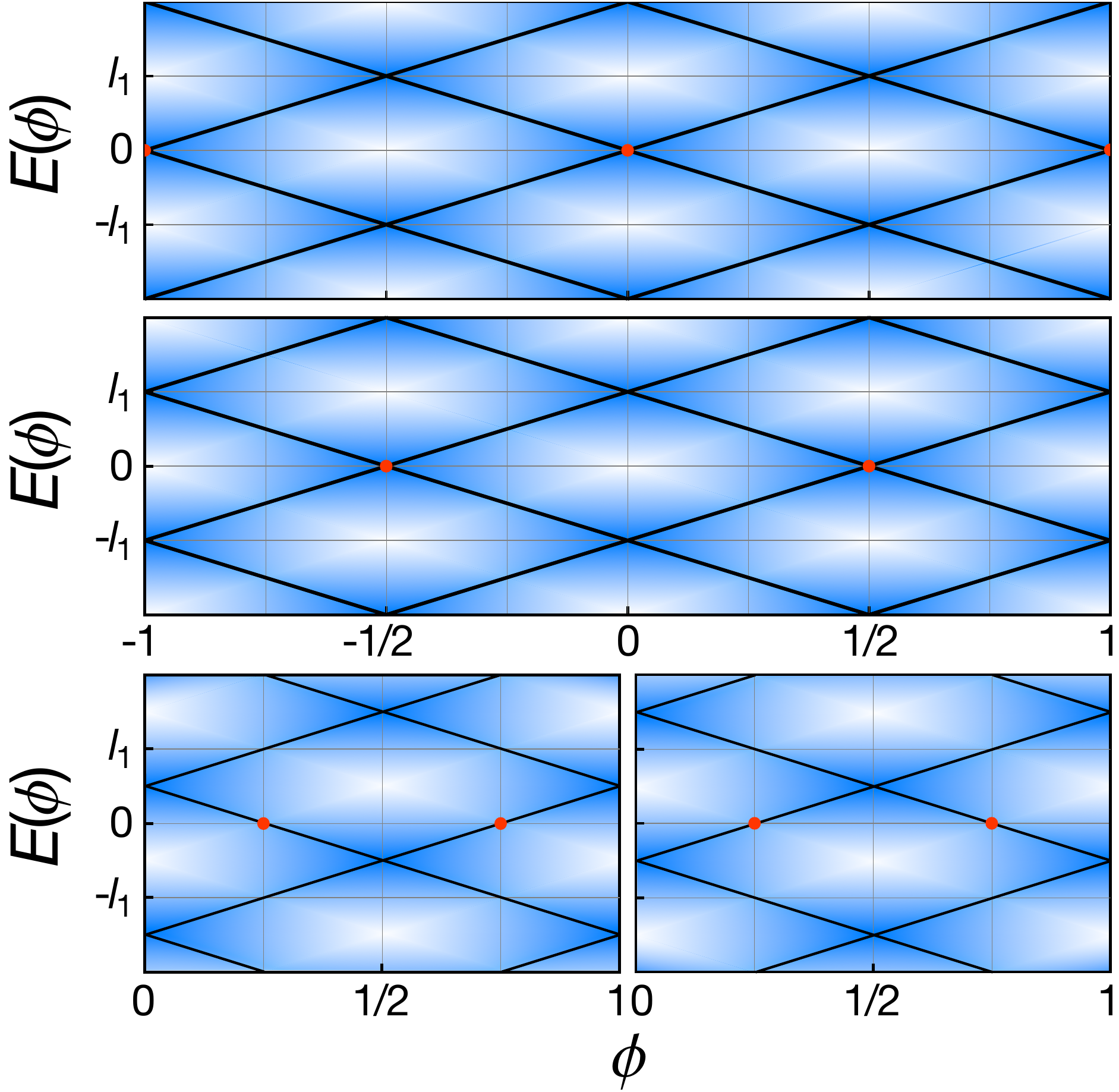}
\put(-6,94.5){({\bf a})}
\put(-6,66.5){({\bf b})}
\put(-6,34.5){({\bf c})}
\end{overpic}
\caption{The energy spectrum of a cylinder in the normal state depends on the numbers $N$ and $M$, which parametrize the circumference and height of the cylinder. The black lines represent the energy levels for a 1D ring with $M=1$ and (a) $N/4$ an integer, (b) $N/4$ a half integer and (c) $N$ an odd number. In (a) and (b), level crossings occur at each multiple of the maximum Doppler shift for $\phi=1/2$, denoted by $l_1$ (see section~\ref{sec3}). For odd $N$, two different spectra without level crossings at $E(\phi)=0$ are possible [$N=4n+1$ (left) and $N=4n-1$ (right) with $n\in\mathbb{N}$]. At the red points, a level crosses the Fermi energy $E_F=0$. For $M\gg1$, the levels split up and form a quasi continuous DOS that depends on the ratio $N/M$ (blue patches).}
\label{Fig0.1}
\end{figure}

The properties of a finite-size multiply connected superconductor depend sensitively on the discrete energy spectrum in the normal state, in particular in circular symmetric geometries. To understand the SC spectrum of the discrete $N\times M$ lattice, we therefore have to characterize first its normal state spectrum.
To illustrate the problem, we consider the tight-binding spectrum of a 1D ring with $N$ lattice sites and nearest-neighbor hopping $t$. A half-filled band corresponds to chemical potential $\mu=0$ in equation~(\ref{s1}). If $N/4$ is an integer, there is an energy level at energy $\epsilon=0$ for $\phi=0$, where $\phi=\Phi/\Phi_0$ is the dimensionless magnetic flux. If $N/4$ is a half-integer, the levels are symmetrically distributed above and below $\epsilon=0$ (figure~\ref{Fig0.1}). As a function of flux, the spectrum is $h/e$-periodic in both cases. If $N$ is odd, there are two possible configurations of energy levels, as shown in figure~\ref{Fig0.1}(c). In both configurations with odd $N$, two levels cross $E_F$ in one flux period. The combination of a particle-like and a hole-like band, used to construct the SC spectrum, then becomes $hc/2e$ periodic.
These number dependent, qualitative differences control the flux dependence of the normal persistent current, as was shown by B\"uttiker \etal and by Cheung \etal \cite{Landauer,cheung:88}.

Whenever an energy level crosses $E_F$ with increasing flux, the current reverses its sign, thus it is $h/e$-periodic for even $N$ and either paramagnetic or diamagnetic in the vicinity of $\phi=0$, and it is $hc/2e$-periodic for odd $N$. The lattice-size dependence persists also in rings with electron-electron interactions \cite{fye:91,fye:92,waintal} or in mesoscopic SC islands \cite{mineev8} and in particular in a 2D cylinder geometry with circumference $Na$ and height $Ma$. Each energy level of the 1D case splits up into $M$ levels, which results in a characteristic flux dependence of the spectral density. For special ratios $N/M$, the flux values where the 1D levels cross have a high degeneracy; for $N=M$, the degree of degeneracy is $M$. For the latter case, the differences between the spectrum for integer and half-integer flux values are most pronounced; they are similar to the 1D spectrum of figure~\ref{Fig0.1}(a), if $N$ and $M$ are even, and similar to the spectrum of figure~\ref{Fig0.1}(b), if $N$ and $M$ are odd. For $N=M\pm1$, the spectrum is almost $hc/2e$-periodic, which is the extension of the odd $N$ case in the 1D ring. Away from these special choices of $N$ and $M$, the degeneracies are lifted, indicated by the blue shaded \textquotedblleft patches\textquotedblright\  in figure~\ref{Fig0.1}. The inclusion of a next-nearest neighbor hopping term or a change of $\mu$ in equation~(\ref{s1}) has a similar effect, as shown by Zhu \cite{zhu:08}. The size of the normal persistent current circulating around the cylinder is controlled by the change of the DOS near $E_F$ upon increasing $\phi$. Since normal persistent currents in metallic rings are typically $hc/e$ periodic \cite{AB, washburn:92}, we will choose $N=M$ and $\mu=0$ for our model study, where  the $hc/e$ periodicity of the spectrum is most clearly established, and we will use even $N$ and $M$ for all subsequent calculations. In this section and in section~\ref{sec3} we show how this size dependent features survive into a SC state with gap nodes.

The starting point for our investigations is the BSC-theory formulated on a flux threaded cylinder with circumference $Na=2\pi Ra$ and height $Ma$, where $R$ is the dimensionless radius of the cylinder and $a$ the lattice constant. The pairing Hamiltonian is given by
\begin{equation}
{\cal H}=\sum_{\kv,s}\epsilon_\kv(\phi)c_{\kv s}^\dag c_{\kv s}+\sum_{\kv}\left[\Delta_\kv^*(\qv)c_{\kv\uparrow}c_{-\kv+\qv\downarrow}+\Delta_\kv(\qv)c^\dag_{-\kv+\qv\downarrow}c^\dag_{\kv\uparrow}\right],
\label{s0}
\end{equation}
where $\kv=(\kp,\kz)$ with $\kp=n/R$ and $n\in\{-N/2+1,\dots,N/2\}$. In the $z$-direction along the axis of the cylinder, we choose open boundary conditions, which allow for even-parity solutions with $\kz=(2m_e-1)\pi/M$ and odd-parity solutions with $\kz=2\pi m_o/M$,  where $m_e,m_o\in\{1,\dots,M/2\}$. The operators $c^\dag_{\kv s}$ and $c_{\kv s}$ are creation and annihilation operators for electrons with crystal angular momentum $\hbar n$ and crystal momentum $\hbar \kz/a$.
The eigenenergies of free electrons moving on a discrete lattice on the surface of the flux threaded cylinder have the form
\begin{equation}
\epsilon_\kv(\phi)=-2t\left[\cos\left(\kp-\frac{\phi}{R}\right)+\cos\kz\right]-\mu.
\label{s1}
\end{equation}
For $R\gg 1$, $\epsilon_\kv(\phi)$ can be expanded to linear order in $\phi/R$ and
\begin{equation}
\epsilon_\kv(\phi)-\epsilon_\kv(0)\approx-2t\frac{\phi}{R}\sin\kp.
\label{s1.1}
\end{equation}
is commonly called the Doppler shift. 

The superconducting order parameter in the pairing Hamiltonian (\ref{s0}) is defined through
\begin{equation}
\Delta_\kv(\qv,\phi)\equiv\Delta_q(\phi) g(\kv)=\frac{1}{2}\sum_{\kv'}V(\kv,\kv')\langle c_{\kv'\uparrow}c_{-\kv'+\qv\downarrow}-c_{\kv'\downarrow}c_{-\kv'+\qv\uparrow}\rangle
\label{s01}
\end{equation}
where $V(\kv,\kv')$ is the pairing interaction. Here we choose the interaction in the separable form $V(\kv,\kv')=Vg(\kv)g(\kv')$ with the pairing interaction strength $V$. The order parameter $\Delta_\kv(\qv,\phi)$ represents spin-singlet Cooper pairs with total  crystal angular momentum $\hbar q$. On the cylinder, the coherent motion of the Cooper pairs is possible only in the azimuthal direction, therefore $\qv=(q/R,0)$ with $q\in\{-N/2+1,\dots,N/2\}$. The quantum number $q$ is chosen to minimize the free energy. The $\phi$-dependence of $\Delta_q(\phi)$ enters through the self-consistency condition and has been discussed extensively in \cite{czajka:05} and \cite{loder:08.2} for $s$-wave pairing, where $g(\kv)\equiv{\rm const}$. Since $\Delta_q(\phi)$ varies only little with $\phi$, we shall postpone the discussion of the flux-dependence of the $d$-wave order parameter to the numerical evaluations of section~\ref{sec4} and start our analytical calculation with a $\phi$ and $q$ independent order parameter $\Delta_q(\phi)\equiv\Delta$. As in our preceding work \cite{loder:08.2}, we take $q={\rm floor}(2\phi+1/2)$ in a first step, such that $\phi-q/2$ is $hc/2e$ periodic; eventual deviations from this relation will be discussed in section~\ref{sec4}. Since the Hamiltonian (\ref{s0}) is invariant under the simultaneous transformation $\phi\rightarrow\phi+1$ and $q\rightarrow q+2$, it is sufficient to consider $q=0$ or $1$ and the corresponding flux sectors $-1/4\leq\phi<1/4$ and $1/4\leq\phi<3/4$, respectively.

\begin{figure}[t]
\centering
\begin{overpic}
[height=5cm]{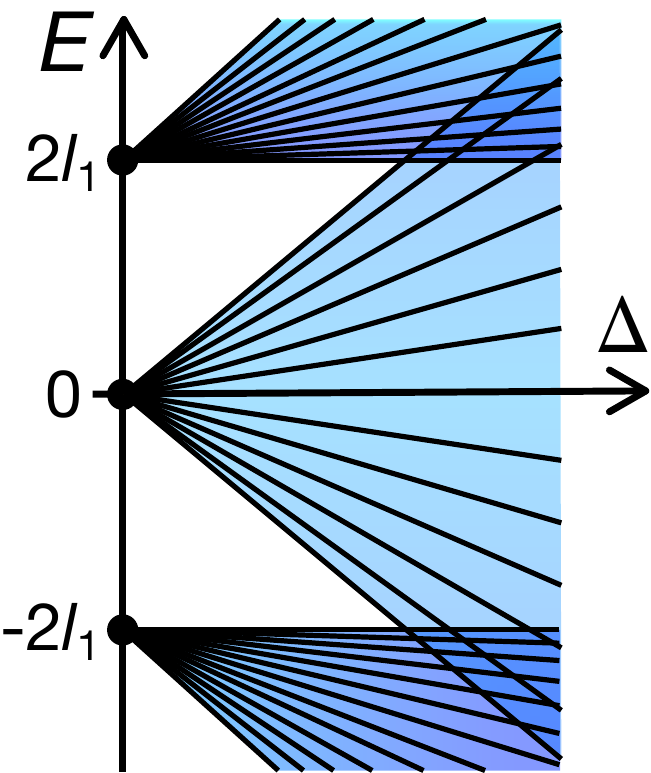}
\put(-26,85){$\begin{array}{l}({\bf a})\\\phi=0,\\q=0\end{array}$}
\end{overpic}
\hspace{2cm}
\begin{overpic}
[height=5cm]{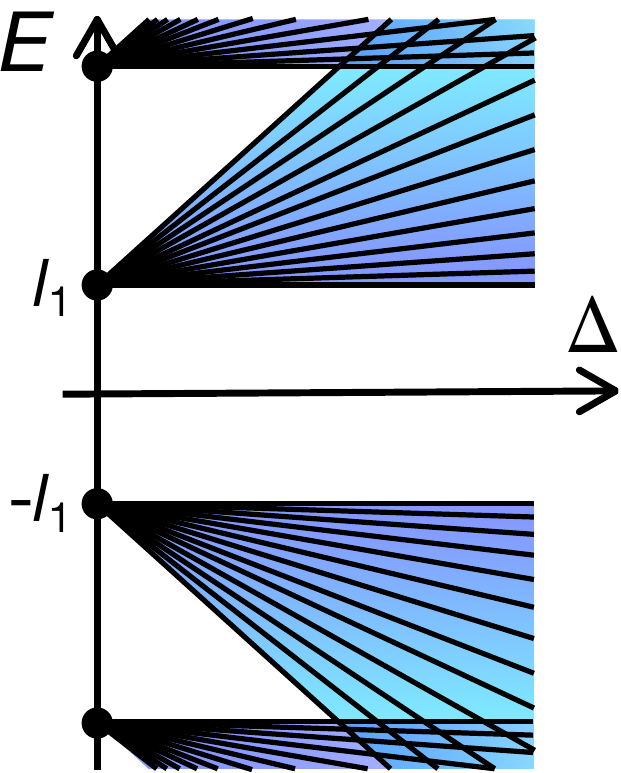}
\put(-30,85){$\begin{array}{l}({\bf b})\\\phi=1/2,\\q=1\end{array}$}
\end{overpic}
\caption{Scheme for the evolution of the multiply degenerate energy levels in the normal state with increasing order parameter $\Delta$ for the $d$-wave SC state. For $\phi=0$ and $q=0$ (a), there is a degenerate level  $E(\kv,0,0)=0$ that splits up for finite $\Delta$ into levels spreading between $-\Delta$ and $\Delta$. For $\phi=1/2$ and $q=1$ (b), there is an energy gap around $E=0$ of width $2l_1$, which persists into the SC state.}
\label{Fig0.2}
\end{figure}

The diagonalization of the Hamiltonian (\ref{s0}) leads to the quasi-particle dispersion
\begin{equation}
E_\pm(\kv,\qv ,\phi)=\frac{\epsilon_\kv(\phi)-\epsilon_{-\kv+\qv}(\phi)}{2}\pm\sqrt{\Delta_\kv^2+\epsilon^2(\kv,\qv,\phi)},
\label{s02}
\end{equation}
with $\epsilon(\kv,\qv,\phi)=[\epsilon_\kv(\phi)+\epsilon_{-\kv+\qv}(\phi)]/2$. Expanding $E_\pm(\kv,\qv,\phi)$ to linear order in both $\phi/R$ and $q/R$ gives
\begin{equation}
E_\pm(\kv,\qv,\phi)\approx-e_q(\kv)\pm
\sqrt{\Delta_\kv^2+\left(\epsilon_\kv(0)-l_q(\kv)\right)^{2}},
\label{s02.1}
\end{equation}
where
\begin{equation}
e_q(\kv)=\frac{\phi-q/2}{R}2t\sin\kp \\ {\rm and} \\ l_q(\kv)=\frac{tq}{R}\sin\kp.
\end{equation}
In the normal state $\Delta=0$, the additive combination of $e_q(\kv)$ and $l_q(\kv)$ leads to the $\qv$-independent spectrum equation~(\ref{s1}). For $\Delta>0$, the spectrum (\ref{s02.1}) differs for even and odd $q$, except for special ratios of $N$ and $M$, as discussed above. This difference is crucial for nodal SC states, as shown schematically in figure~\ref{Fig0.2} (and especially for $d$-wave pairing in figure~\ref{Fig7}): The condition $\kp\approx\kz$ for levels close to $E_F$ causes a level spacing $\delta_F\approx2l_1(\kv_F)$ for small $\Delta$, where $\kv_F$ is the Fermi momentum. For $N$ and $M$ even and $q=0$, the degenerate energy level at $E=E_F=0$ splits into $M$ levels for increasing $\Delta$, which spread between $-\Delta$ and $\Delta$. For $q=1$, the degenerate levels closest to $E_F$ are located at $E=\pm |l_1(\kv_F)|$, thus a gap of $2l_1(\kv_F)$ remains in the SC spectrum. If $N$ and $M$ are odd, the spectra for even and odd $q$ (figure~\ref{Fig0.2}(a) and (b)) are exchanged, and if either $N$ or $M$ is odd, the spectrum is a superposition of (a) and (b).

The gauge invariant circulating supercurrent is given by
\begin{equation}
J_\qv(\phi)=\frac{e}{h}\sum_{\kv,s}v_\kv n_{\kv s}(\qv),
\label{s13}
\end{equation}
where $v_\kv={\partial\epsilon_\kv(\phi)}/{\partial(R\kp)}$ is the group velocity of the single-particle state with eigenenergy $\epsilon_\kv(\phi)$.
The spin independent occupation probability of this state is
\begin{equation}
\fl n_{\kv s}(\qv)=\langle c_{\kv s}^\dag c_{\kv s}\rangle(\qv)
=u^2(\kv,\qv,\phi)f(E_+(\kv,\qv,\phi))-v^2(\kv,\qv,\phi)f(E_-(\kv,\qv,\phi)),
\label{s14}
\end{equation}
where $f(E)=1/(1+e^{E/k_{\rm B}T})$ is the Fermi distribution function for the temperature $T$. The Bogoliubov amplitudes are
\begin{equation}
u^2(\kv,\qv,\phi)=\frac{1}{2}\left[\frac{\epsilon(\kv,\qv,\phi)}{E(\kv,\qv,\phi)}+1\right]\\{\rm and}\\
v^2(\kv,\qv,\phi)=1-u^2(\kv,\qv,\phi).
\end{equation}

From equations~(\ref{s13}) and (\ref{s14}), the supercurrent in the cylinder is obtained by evaluating either the sum numerically, as discussed in section~\ref{sec4}, or from the approximative analytic solution in section~\ref{sec3}, which allows insight into the origin of the $hc/e$-periodicity in nodal superconductors.
\section{Analytic Solution and Qualitative Discussion} \label{sec3}

An analytic evaluation of the supercurrent is possible only in the thermodynamic limit where the sum over discrete eigenstates is replaced by an integral. For a multiply connected geometry, this limit is not properly defined because the supercurrent or the Doppler shift vanish in the limit $R\rightarrow\infty$. Care is needed to modify the limiting procedure in a suitable way to access the limit of a large but non-infinite radius of the cylinder. In this limit it is mandatory to consider the supercurrent density $j_\qv(\phi)=J_\qv(\phi)/M$ rather than the supercurrent $J_\qv(\phi)$.
In this scheme, we treat the DOS as a continuous function in any energy range where the level spacing is $\propto1/NM$, but we keep the finite energy gap of width $2l_q(\kv_F)\propto1/R\propto1/N$ around $E_F$ in the odd-$q$ sectors. 

For a tight binding energy spectrum as defined in equation~(\ref{s1}), the density of states is a complete elliptic integral of the first kind. For the purpose of an analytic calculation, a quadratic spectrum with a constant density of states in two dimensions is a more appropriate starting point. We use the expanded form of equation~(\ref{s1}):
\begin{equation}
\epsilon_\kv(\phi)=t\left[\left(\kp-\frac{\phi}{R}\right)^2+\kz^2\right]-\mu',
\label{s1.1}
\end{equation}
where $\mu'=\mu+4t$. The quadratic spectrum (\ref{s1.1}) has no upper bound and the sum in equation~(\ref{s13}) correspondingly extends from $-\infty$ to $\infty$ for both $\kp$ and $\kz$.

Some algebraic steps are needed to rearrange the sum in equation~(\ref{s13}) suitably to convert it into an integral.
For finite $\phi$, $\epsilon_\kv(\phi)\neq\epsilon_{-\kv}(\phi)$, and consequently the sum has to be decomposed into a component with $\kp\geq0$ and a second one with $\kp<0$. We therefore take $\kp\geq0$ and write $v_\kv$ as
\begin{equation}
v_{\pm\kv}=\frac{2t}{R}\left(\pm\kp-\frac{\phi}{R}\right)=v_d(\kv)\pm v_p(\kv),
\label{s242}
\end{equation}
where $v_d(\kv)=-2t\phi/R^2$ is the diamagnetic contribution and $v_p(\kv)=2t\kp/R$,  a paramagnetic contribution, respectively \cite{scalapino:93}.

In a continuous energy integration, the Doppler shift is noticeable only in the vicinity of $E_F$. On the Fermi surface, $\kp$ and $\kz$ are related by:
\begin{equation}
k_{\varphi,F}(\kz)=\sqrt{\frac{\mu'}{t}-\kz^2}.
\label{s244}
\end{equation}
In this spirit we approximate $e_q(\kv)$ and $l_q(\kv)$ by $
e_q(\kz)\approx2t(\phi-q/2)k_{\varphi,F}(\kz)/R$ and  $l_q(\kz)\approx tq k_{\varphi,F}(\kz)/R$, respectively.
The eigenenergies (\ref{s02.1}) near $E_F$ are thereby rewritten as
\begin{eqnarray}
E_+(\pm\kp,\kz,\qv,\phi)=\mp e_q(\kz)+\sqrt{\Delta_\kv^2+\left(\epsilon_\kv(0)\mp l_q(\kz)\right)^2}\\
E_-(\pm\kp,\kz,\qv,\phi)=\mp e_q(\kz)-\sqrt{\Delta_\kv^2+\left(\epsilon_\kv(0)\mp l_q(\kz)\right)^2}
\label{s02.2}
\end{eqnarray}

For the evaluation of the supercurrent $J_\qv(\phi)$ in equation~(\ref{s13}), the sum over $\kv$ is now replaced by an integral over $\kp$ and $\kz$, which is then performed by integrating over the normal state energy $\epsilon$ and an angular variable $\theta$. According to our scheme for replacement of discrete energy levels by a continuous spectrum, the DOS becomes gapless in the limit $M\rightarrow\infty$ for $q=0$, although $N$ is kept finite. For $q=1$ instead, a $\kz$-dependent gap $2|l_1(\kz)|$ remains. Thus we replace $\epsilon_\kv(0)\mp |l_q(k_z)|$ by the continuous quantity $\epsilon\pm|l_q(E_F,\theta)|$. 
In summary, the procedure is defined by the following steps:
\begin{eqnarray}
\fl \sum_{\kv}
\overset{R,M\rightarrow\infty}{\longrightarrow}\frac{RM}{2\pi}\int_{0}^\infty{\rm d}\kp{\rm d}\kz
=\frac{RM}{2\pi}\int_0^\infty{\rm d}kk\int_{-\pi/2}^{\pi/2}{\rm d}\theta
=M{\cal N\!}\int_{-\mu'}^{\mu'}{\rm d}\epsilon\int_{-\pi/2}^{\pi/2}{\rm d}\theta,
\label{s245}
\end{eqnarray}
where we use the parametrization
\begin{equation}\left(
\begin{array}{l}\kp\\\kz\end{array}\right)=\left(\begin{array}{l}k\cos\theta\\ k\sin\theta \end{array}\right)=\sqrt{\frac{\epsilon+\mu'}{t}}\left(\begin{array}{l}\cos\theta\\\sin\theta\end{array}\right),
\label{s246}
\end{equation}
with $\epsilon=tk^2-\mu'$ and where ${\cal N}=R/4\pi t$ is the constant DOS in the normal state. The energy integral runs over the whole tight-binding band width $8t$ with the Fermi energy $E_F=0$ in the center of the band. Correspondingly, we integrate from $-\mu'$ to $\mu'$. Furthermore, the Doppler shift is parametrized for $\epsilon\approx E_F$ as
\begin{equation}
e_q(\theta)=\frac{\phi-q/2}{R}2t\sqrt{\mu'/t}\cos\theta\\{\rm and}\\ l_q(\theta)=\frac{tq}{R}\sqrt{\mu'/t}\cos\theta,
\end{equation}
where the function $l_q(\theta)$ is positive for all allowed values of $\theta$.
The supercurrent thus becomes
\begin{eqnarray}
\fl j_\qv(\phi)=\frac{1}{M}\frac{e}{h}\left[\sum_{\kp>0,\kz,s}v_\kv n_{\kv s}(\qv)+\sum_{\kp<0,\kz,s}v_\kv n_{\kv s}(\qv)\right]\nonumber\\
\approx2{\cal N}\frac{e}{h}\!\int_{-\pi/2}^{\pi/2}\!\!{\rm d}\theta\int_{-\mu'}^{\mu'}\!{\rm d}\epsilon[n_{q+}(\epsilon,\theta)v_{+}(\epsilon,\theta)\label{s29}+n_{q-}(\epsilon,\theta)v_{-}(\epsilon,\theta)],
\label{s247}
\end{eqnarray}
where $n_{q\pm}(\epsilon,\theta)= n_{\pm\kv(\epsilon,\theta)}(\qv)$ and $v_\pm(\epsilon,\theta)=v_{\pm\kv(\epsilon,\theta)}$. The factor 2 in equation~(\ref{s247}) originates from the spin sum.
We collect the terms proportional to $v_d(\epsilon,\theta)=-2t\phi/R^2$ into a diamagnetic current contribution $j_d$ and the terms proportional to $v_p(\epsilon,\theta)=2tk_{\varphi,F}(\epsilon,\theta)/R$ into a paramagnetic contribution $j_p$. Using $f(-E)=1-f(E)$, equation (\ref{s247}) simplifies to
\begin{equation}
\fl j_d=4{\cal N}\frac{e}{h}\!\int_{-\pi/2}^{\pi/2}{\rm d}\theta\int_{l_q(\theta)}^{\mu'}{\rm d}\epsilon\, v_d\left(\epsilon,\theta\right)\frac{\epsilon}{\sqrt{\Delta^2+\epsilon^2}}
\left[f(E+e_q(\theta))-f(-E+e_q(\theta))\right],
\label{s81.0}
\end{equation}
\begin{equation}
\fl j_p=4{\cal N}\frac{e}{h}\!\int_{-\pi/2}^{\pi/2}{\rm d}\theta\int_{l_q(\theta)}^{\mu'}{\rm d}\epsilon\, v_p\left(\epsilon,\theta\right)
\left[f(-E-e_q(\theta))-f(-E+e_q(\theta))\right],
\label{s82.0}
\end{equation}
where $j_d=j_d(q,\phi)$ and $j_p=j_p(q,\phi)$.
Here, the integration is over positive values of $\epsilon$ only and the lower boundaries of the integration over $\epsilon$ are controlled by $l_q(\theta)$. Since $l_q(\kv)=0$ at the minimum of the band ($\epsilon=-\mu'$), the upper integral boundary remains $\mu'$. We used the abbreviations  $\Delta=\Delta(\theta)$ and $E=E(\epsilon,\theta)=\sqrt{\Delta^2(\theta)+\epsilon^2}$. The current $j_d$ turns out to be diamagnetic in the even-$q$ flux sectors and paramagnetic in the odd-$q$ sectors. For even $q$, it is equivalent to the diamagnetic current obtained from the London equations \cite{pethick:79,tinkham3}. The current $j_p$ has always the inverse sign of $j_d$ and is related to the quasi-particle current as shown below. As presented in section \ref{sec2}, $E$ displays a distinct spectra in the even-$q$ and odd-$q$ flux sectors. To analyze the flux dependent properties of the spectra and the current, we distinguish the case of $s$-wave pairing (or any other superconducting state with a complete energy gap) and the case of unconventional pairing with nodes in the gap function. For the latter, we focus on $d$-wave pairing.
\subsection{$\bm s$-Wave Pairing Symmetry}
For $s$-wave pairing, $\Delta(\epsilon,\theta)\equiv\Delta$ is constant. Therefore, if we assume that $\Delta\geq e_q(\theta)$ for all $\theta$, the lower energy integration boundaries in equations~(\ref{s81.0}) and (\ref{s82.0}) are equal to $\Delta$. Thus $j_\qv(\phi)$ is equal in both the even-$q$ and odd-$q$ flux sectors and the flux periodicity is $hc/2e$. However, if $\Delta<\max_\theta e_q(\theta)$, equation~(\ref{s13}) has to be evaluated exactly, the procedure and results of which have been presented in \cite{loder:08.2}. 

With $\epsilon=\sqrt{E^2-\Delta^2}$, equations~(\ref{s81.0}) and (\ref{s82.0}) transform into integrals over $E$ with ${\rm d}\epsilon=D_s(E)\,{\rm d}E$, where
\begin{equation}
D_s(E)=\frac{\partial\epsilon}{\partial E}=\left\{\begin{array}{lll}E\,(E^2-\Delta^2)^{-1/2}&{\rm for}&E\geq\Delta\\0&{\rm for}&E<\Delta\end{array}\right.
\label{s6}
\end{equation}
is the SC density of states for $s$-wave pairing.
This leads to
\begin{equation}
\fl j_d=4{\cal N}\frac{e}{h}\!\int_{-\pi/2}^{\pi/2}{\rm d}\theta\int_\Delta^{\mu'}{\rm d}E v_d\left(\sqrt{E^2-\Delta^2},\theta\right)\left[f(E+e_q(\theta))-f(-E+e_q(\theta))\right],
\label{s81}
\end{equation}
\begin{equation}
\fl j_p=4{\cal N}\frac{e}{h}\!\int_{-\pi/2}^{\pi/2}d\theta\int_\Delta^{\mu'}dED_s(E)v_p\left(\sqrt{E^2-\Delta^2},\theta\right)\left[f(-E-e_q(\theta))-f(-E+e_q(\theta))\right].
\label{s82}
\end{equation}
At $T=0$, we find
\begin{eqnarray}
\fl j_d=-4{\cal N}\frac{e}{h}\!\int_{-\pi/2}^{\pi/2}{\rm d}\theta\int_\Delta^{\mu'}{\rm d}E\,2t\frac{\phi-q/2}{R^2}=-2(\mu'-\Delta)\frac{e}{h} \frac{\phi-q/2}{R},
\label{s9}\\[3mm]
\fl j_p=4{\cal N}\frac{e}{h}\!\int_{-\pi/2}^{\pi/2}d\theta\int_\Delta^{e_q(\theta)}dED_s(E)\frac{2t}{R}\sqrt{\frac{\epsilon+\mu'}{t}}\cos\theta\nonumber\\
=\frac{8t{\cal N}}{R}\frac{e}{h}\sqrt{\frac{\mu'}{t}}\int_{-\pi/2}^{\pi/2}d\theta\cos\theta\int_\Delta^{e_q(\theta)}dED_s(E)+{\cal O}\left(\frac{\epsilon}{t}\right)^2.
\label{s10}
\end{eqnarray}
In the integral over of $j_p$, the inequality $\epsilon/t\ll1$ applies, and terms of order ${\cal O}(\epsilon/t)^2$ are negligible. 

The current $j_d$ becomes independent of the SC density of states. Its size is essentially proportional to $E_F$, as long as $\mu'\gg\Delta$ holds. The paramagnetic current $j_p$ depends on the absolute value of the order parameter and on its symmetry.

If $\Delta>e_q(\theta)$ for all values of $\theta$, then $j_p=0$ and the supercurrent $j_\qv=j_d$ consists of the diamagnetic part only. For $T>0$, $j_d$ decreases slightly, but remains of the same order of magnitude. The current $j_p$ increases with increasing $T$ and reaches its maximum value at $T_c$. For finite temperatures $j_p$ is usually denoted as the quasi-particle current. The entire supercurrent is always the sum of the diamagnetic current $j_d$ and the quasi-particle current $j_p$,  and therefore decreases with temperature and vanishes at $T_c$ \cite{vonoppen:92}. The quasi-particle current has the same flux periodicity as the supercurrent, even though it is carried by single quasiparticle excitations. In the normal state ($\Delta=0$),
\begin{eqnarray}
\fl j_p=\frac{8t{\cal N}}{R}\frac{e}{h}\sqrt{\frac{\mu'}{t}}\int_{-\pi/2}^{\pi/2}d\theta\cos\theta\int_0^{e_q(\theta)}dE=4\mu'\frac{e}{h} \frac{\phi-q/2}{R\pi}\int_{-\pi/2}^{\pi/2}d\theta\cos^2\theta\nonumber\\
=2\mu'\frac{e}{h}\frac{\phi-q/2}{R},
\label{s52}
\end{eqnarray}
which cancels $j_d$ exactly in the limit $M\rightarrow\infty$. \footnote{In this procedure, the normal persistent current vanishes, but this is unproblematic here because the normal current above $T_c$ is exponentially small for $k_{\rm B}T_c\gg\delta_F$.}
\subsection{Unconventional Pairing with Gap Nodes}
\begin{figure}[t]
\centering
\begin{overpic}
[width=0.49\columnwidth]{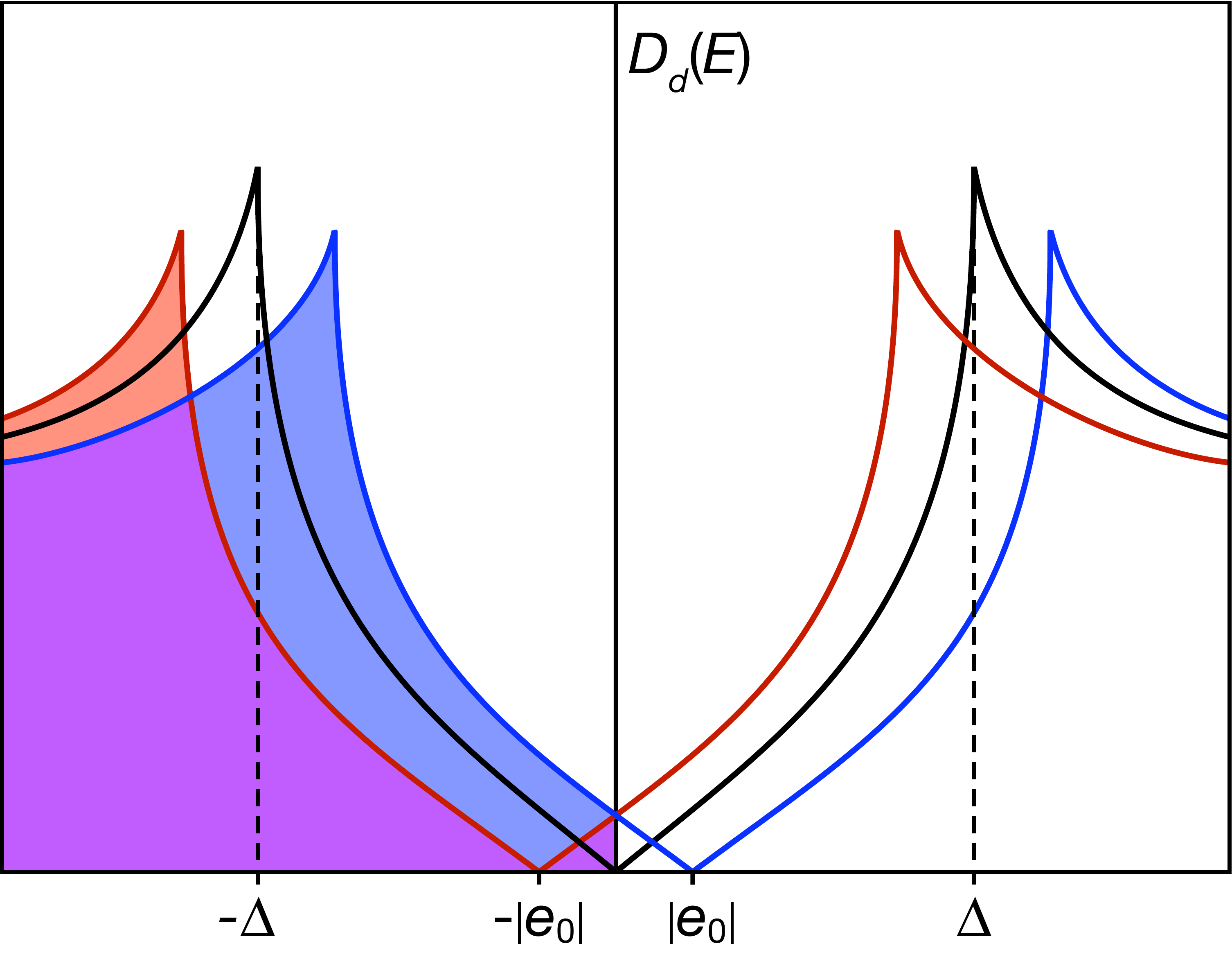}
\put(1,74){({\bf a})}
\end{overpic}
\begin{overpic}
[width=0.49\columnwidth]{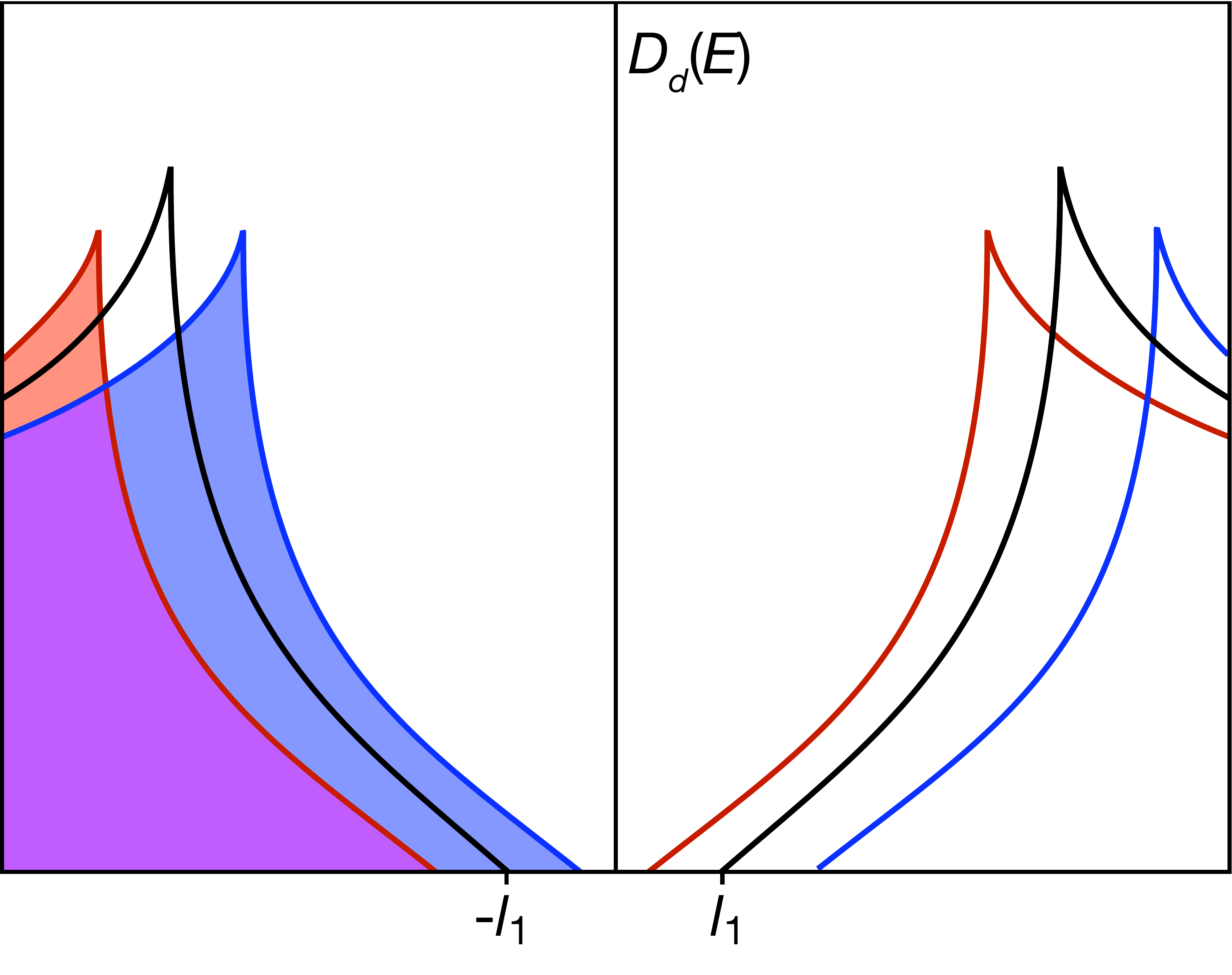}
\put(1,74){({\bf b})}
\end{overpic}
\caption{Scheme for the density of states of a $d$-wave superconductor for $\phi=1/4$, where $e_q(1/4)=l_1/2$. The center of mass angular angular momentum $\hbar q$ of the Cooper pairs is (a) $q=0$ and (b) $q=1$. The energies are Doppler shifted to higher (red) or lower energies (blue). This results in a double-peak structure and for $q=0$ in an overlap of the upper and lower \textquotedblleft band\textquotedblleft\ in the region $-|e_0|<E<|e_0|$ \cite{khavkine:04} and states in the upper band become partially occupied. For $q=1$ there is a gap $l_1$ of the size of the maximum Doppler shift at $\phi=1/4$. The black line represents the density of states (a) for $\phi=0$ and (b) for $\phi=1/2$.}
\label{Fig1}
\end{figure}

For a more general order parameter $\Delta(\theta)$, an analytic solution of equations~(\ref{s81.0}) and (\ref{s82.0}) is hard to obtain. For $s$-wave symmetry, $j_d$ depends only weakly $\Delta$; $j_d$ is indeed maximal for $\Delta=0$. Equation~(\ref{s9}) for $j_d$ is valid also for unconventional order parameter symmetries. Physically, $j_d$ reflects the difference in the DOS of quasi-particle states with orbital magnetic moment parallel and anti-parallel to the external magnetic field. The first group of states is Doppler shifted to lower energies, whereas the latter is Doppler shifted to higher energies. This is schematically shown in figure~\ref{Fig1} for $d$-wave pairing (c.f. \cite{khavkine:04}). In this picture, $j_d$ is proportional to the difference between the area beneath the red and and blue curves representing the DOS arising of bands $E_-(\pm|\kv|,\qv,\phi)<0$ (underlaid red and blue). Therefore we approximate $j_d$ for $\Delta(\theta)\ll\mu'=E_F+4t$ by
\begin{equation}
j_d=-2\mu'\frac{e}{h} \frac{\phi-q/2}{R},
\end{equation}
as given by equation (\ref{s9}) with $\Delta=0$.
On the other hand, $j_p$ is represented by the occupied quasi-particle states in the overlap region of $E_+(\kv,\qv,\phi)$ and $E_-(\kv,\qv,\phi)$ with width $2e_q(\kv_F)$. It is therefore strongly dependent on the characteristic DOS in the vicinity of $E_F$. In figure~\ref{Fig1}(a), which refers to even $q$, the current $j_p$ is determined by the small triangular patch where the upper and lower bands overlap. For odd $q$, the two bands do not overlap, therefore $j_p=0$.

We will now analyze such a scenario for $d$-wave pairing.
With an order parameter $\Delta_\kv=\Delta(\kp^2-\kz^2)\approx\Delta\cos2\theta$. Again, we assume $\Delta>e_q(\theta)$ for all $\theta$; then the integral in equation~(\ref{s82.0}) contains only the nodal states closest to $E_F$, for which the $d$-wave symmetry demands $\kp\approx\kz$. Jointly with equation~(\ref{s244}) this condition fixes the Doppler shift at $E_F$ to the $\kv$-independent value $e_q=(\phi-q/2)\sqrt{2t\mu'}/R$ and $l_q=(q/R)\sqrt{t\mu'/2}$. With the density of states in the $d$-wave superconducting state
\begin{equation}
D_d(E)=\frac{1}{\sqrt{E^2-\Delta^2\cos^22\theta}},
\label{s54}
\end{equation}
equation~(\ref{s82.0}) for the paramagnetic current $j_p$ at $T=0$ then takes the form
\begin{equation}
j_p=4{\cal N}\frac{e}{h}\!\int_{l_q}^{e_q}{\rm d}E\int_{-\pi/2}^{\pi/2}{\rm d}\theta D_d(E)\frac{2t}{R}\sqrt{\frac{\epsilon+\mu'}{t}}\sin\theta.
\end{equation}
In the odd-$q$ flux sectors, $l_q\geq e_q$ for all values of $\phi$, therefore $j_p=0$. In the $q=0$ sector, $l_q=0$ and 
\begin{eqnarray}
\fl j_p\approx\frac{2e}{h\pi}\sqrt{\frac{\mu'}{t}}\int_{0}^{e_q}dE\int_{-\pi/2}^{\pi/2}d\theta\sin\theta\frac{1}{\sqrt{E^2-\Delta^2\cos^22\theta}}
\approx\frac{2e}{\pi h}\sqrt{\frac{\mu'}{t}}\int_{0}^{e_q}dE\frac{E}{\Delta}
=\frac{e}{\pi h\Delta}\sqrt{\frac{\mu'}{t}}e_q^2\nonumber\\
=\frac{2}{\pi\Delta}\sqrt{t\mu'^3}\frac{e}{h}\left(\frac{\phi-q/2}{R}\right)^2,
\label{s55}
\end{eqnarray}
where the same approximations as in the $s$-wave case are applied. 
The dominant contribution to the angular integral ocer $\theta$ originates from the nodal parts, where the integrand can be linearized in $\theta$, such that the integral can be performed approximately (see e.g. \cite{mineev17}).

In the even-$q$ sectors, the total current $j_\qv=j_d+j_p$ finally becomes
\begin{equation}
j_\qv(\phi)=-2\mu'\frac{e}{h}\frac{\phi}{R}\left[1-\frac{\sqrt{t\mu'}}{\pi\Delta}\frac{\phi}{R}\right],
\label{s34}
\end{equation}
which results in the ratio 
\begin{equation}
\frac{j_p}{j_d}=\frac{\sqrt{t\mu'}}{\pi\Delta}\frac{\phi}{R}\equiv b\phi
\label{s45}
\end{equation}
of the two current components.

\begin{figure}[t]
\centering
\begin{overpic}
[width=0.49\columnwidth]{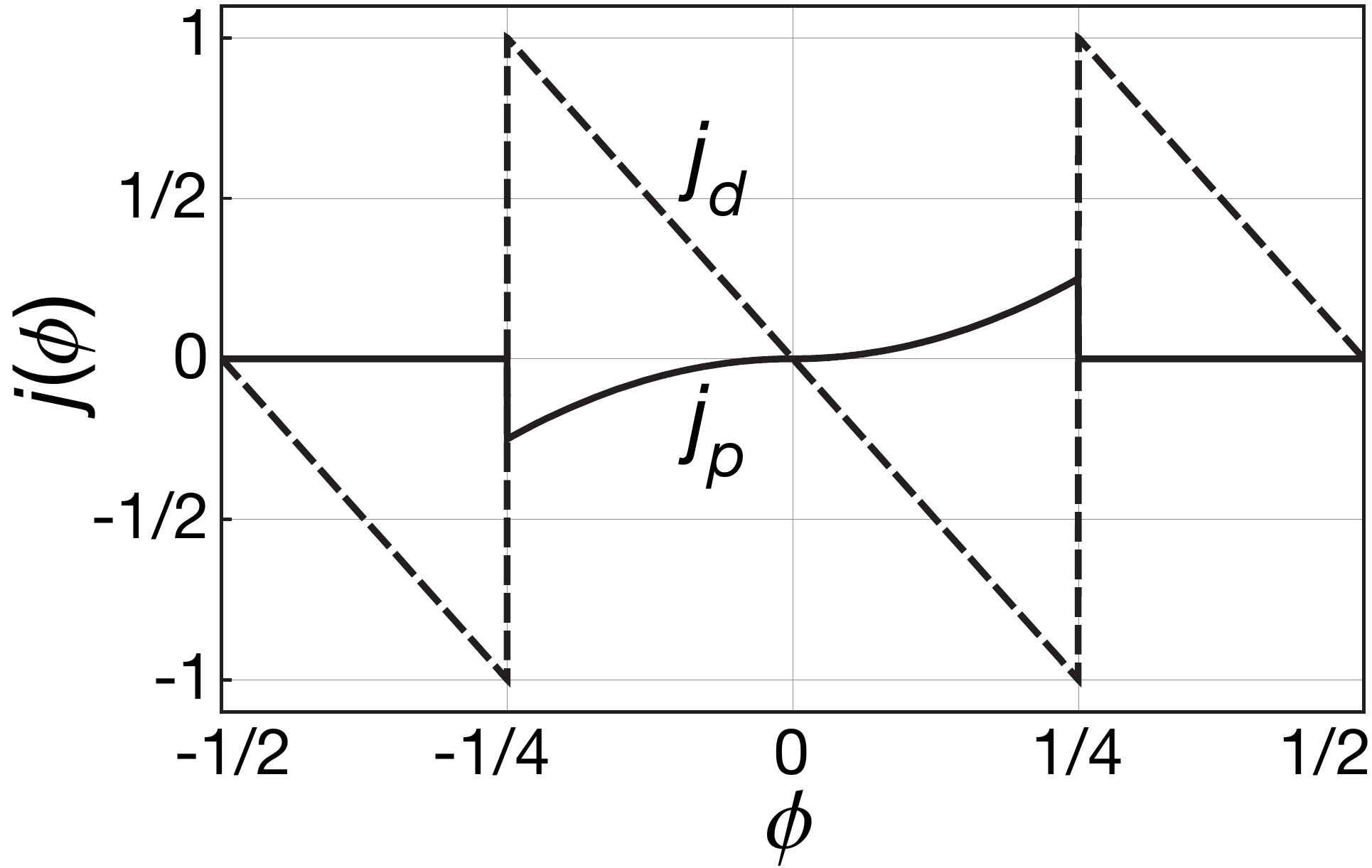}
\put(-0,59){({\bf a})}
\end{overpic}
\begin{overpic}
[width=0.49\columnwidth]{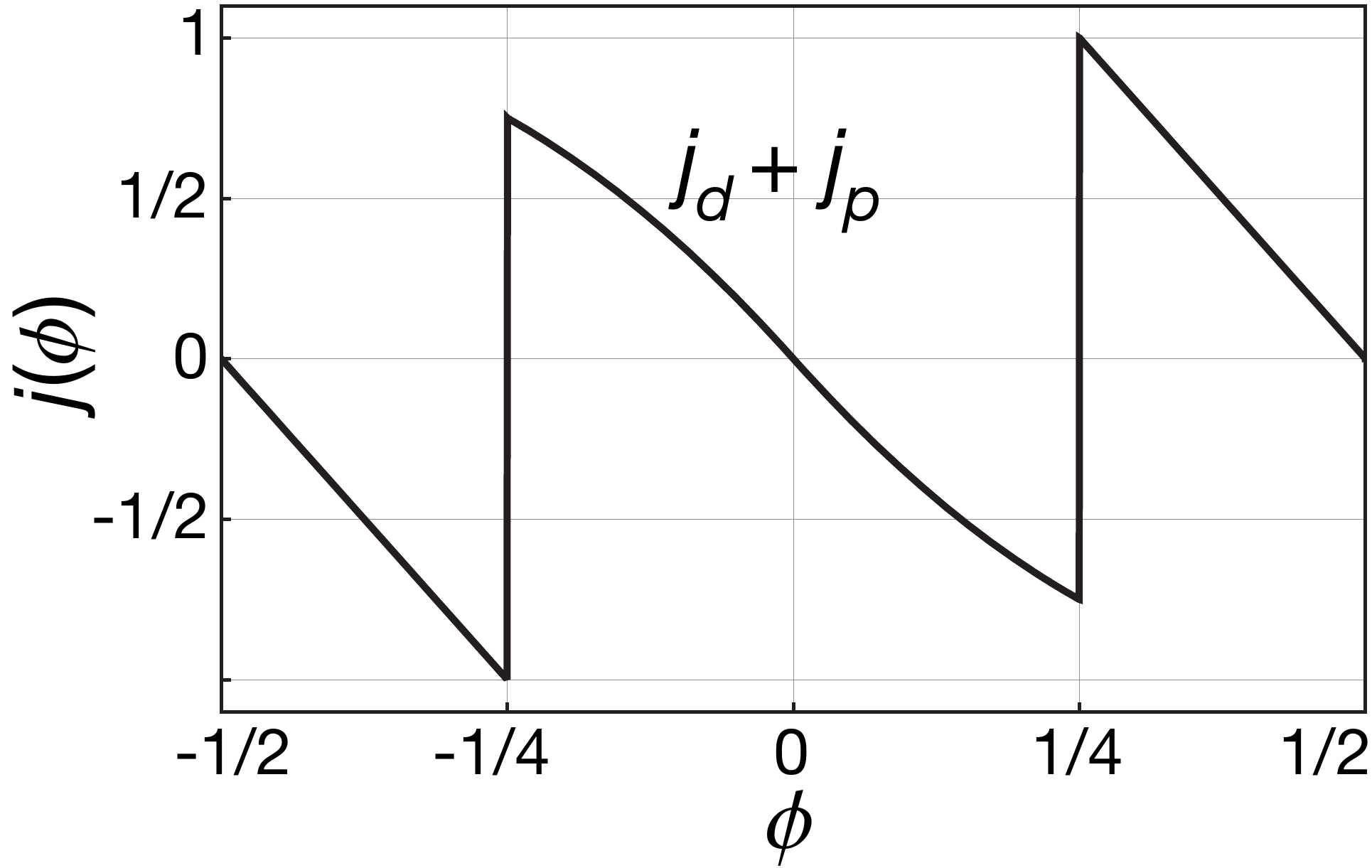}
\put(-0,59){({\bf b})}
\end{overpic}
\caption{The supercurrent density $j_q(\phi)=j_d+j_p$ in a thin $d$-wave cylinder as a function of flux $\phi$ (arbitrary units). Shown is the result of the analytical model (equation~(\ref{s48}))
for the characteristic value $b=0.4$. For $-1/4 < \phi<1/4$, where $q=0$, the current is reduced by a contribution proportional to $\phi^2$, whereas it is linear in $\phi$ otherwise. This gives rise to an overall flux periodicity of $hc/e$.}
\label{Fig2}
\end{figure}

In the odd-$q$ flux sectors, $j_p=0$ and the supercurrent is $j_\qv(\phi)=j_d$. As a function of $\phi$, $j_\qv(\phi)$ is consequently $hc/e$ periodic; within one flux period from $-1/2$ to $1/2$ we represent it as
\begin{equation}
j(\phi)=
-2\frac{\mu'}{R}\frac{e}{h}\left\{\begin{array}{llrl} 
\phi+1/2&{\rm for}&-1/2\leq&\hspace{-2mm}\phi<-1/4,\\ 
\phi(1-b\phi)&{\rm for}& -1/4\leq&\hspace{-2mm}\phi<1/4,\\ 
\phi-1/2&{\rm for}& 1/4\leq&\hspace{-2mm}\phi<1/2, 
\end{array}\right. 
\label{s48}
\end{equation}
(c.f. figure~\ref{Fig2}). The amount by which the supercurrent differs in the even-$q$ and odd-$q$ flux sectors is represented best in the form of Fourier components: the $n$-th Fourier component of $j(\phi)$ is $j_n=\int_{-1/2}^{1/2}d\phi\;j(\phi)e^{2\pi in\phi}$. Here, we denote the first Fourier component by $j_{h/e}$, and the second Fourier component by $j_{h/2e}$ and obtain
\begin{equation}
j_{h/e}
=-2\frac{\mu'}{R}\frac{e}{h}b\frac{8-\pi^2}{16\pi^3}\\{\rm and}\\j_{h/2e}
=-2\frac{\mu'}{R}\frac{e}{h}\frac{4\pi i-b}{16\pi^2}.
\label{s49}
\end{equation}
To leading order in $1/R$, the ratio of the $h/e$ and the $h/2e$ Fourier component therefore is
\begin{equation}
\left|\frac{j_{h/e}}{j_{h/2e}}\right|=\frac{\pi^2-8}{4\pi^2}\frac{\sqrt{2t\mu'}}{\Delta R}\overset{\mu=0}{\longrightarrow}\ \approx0.07\frac{2t}{\Delta R}
\label{s51}
\end{equation}
and scales with the inverse ring diameter. This $1/R$ law is the direct consequence of the $d$-wave density of states $D_d(E)\propto E$. For some other unconventional superconducting states with $D(E)\propto E^n$ in the vicinity of $E_F$, the decay of the $j_{h/e}$ Fourier component results in a $1/R^n$ law. Using equation~(\ref{s51}) to estimate this ratio for a mesoscopic cylinder with a circumference $Ra=2600a\approx$ 1 $\mu$m and a ratio $\Delta/t=0.01$, we obtain $j_{h/e}/j_{h/2e}\approx0.03$.
\section{Numerical Solution for $\bm d$-Wave Pairing at $\bm{T=0}$} \label{sec4}
In this section we evaluate numerically the supercurrent in equation~(\ref{s13}) together with the self-consistency condition
\begin{equation}
\fl \frac{1}{V}=\frac{1}{NM}\sum_{\kv'}\frac{g_\kv^2(q)}{2\sqrt{\Delta_q^2(\phi)g_\kv^2(q)+\epsilon^2(\kv',\qv,\phi)}}\left[f(E_-(\kv',\qv,\phi)-f(E_+(\kv',\qv,\phi)\right],
\label{s401}
\end{equation}
where the $d$-wave pairing symmetry follows from by
\begin{equation}
g_\kv(q)=\cos(\kp-q/2)-\cos\kz
\label{s402}
\end{equation}
and the order parameter is 
$\Delta_{\kv}(\qv,\phi)=\Delta_q(\phi)g_\kv(q)$.
Here we take into account the full $q$- and $\phi$-dependence of $\Delta_\kv(\qv,\phi)$. The $q$-dependence of $g_\kv(q)$ is essential to ensure the invariance of the gap equation~(\ref{s401}) under the replacement $\phi\rightarrow\phi+1$ and $q\rightarrow q+2$. At those flux values for which the total energies
\begin{equation}
E_t(q,\phi)=\sum_{\kv,s}\epsilon_\kv(\phi)n_{\kv s}(\qv).
\end{equation}
are equal, $q$ advances to the next integer.
This flux value may deviate from the values $\phi=(2n-1)/4$, for which we fixed the $q$-sector transitions in section~\ref{sec2}. 

\begin{figure}[t]
\centering
\begin{overpic}
[width=0.49\columnwidth]{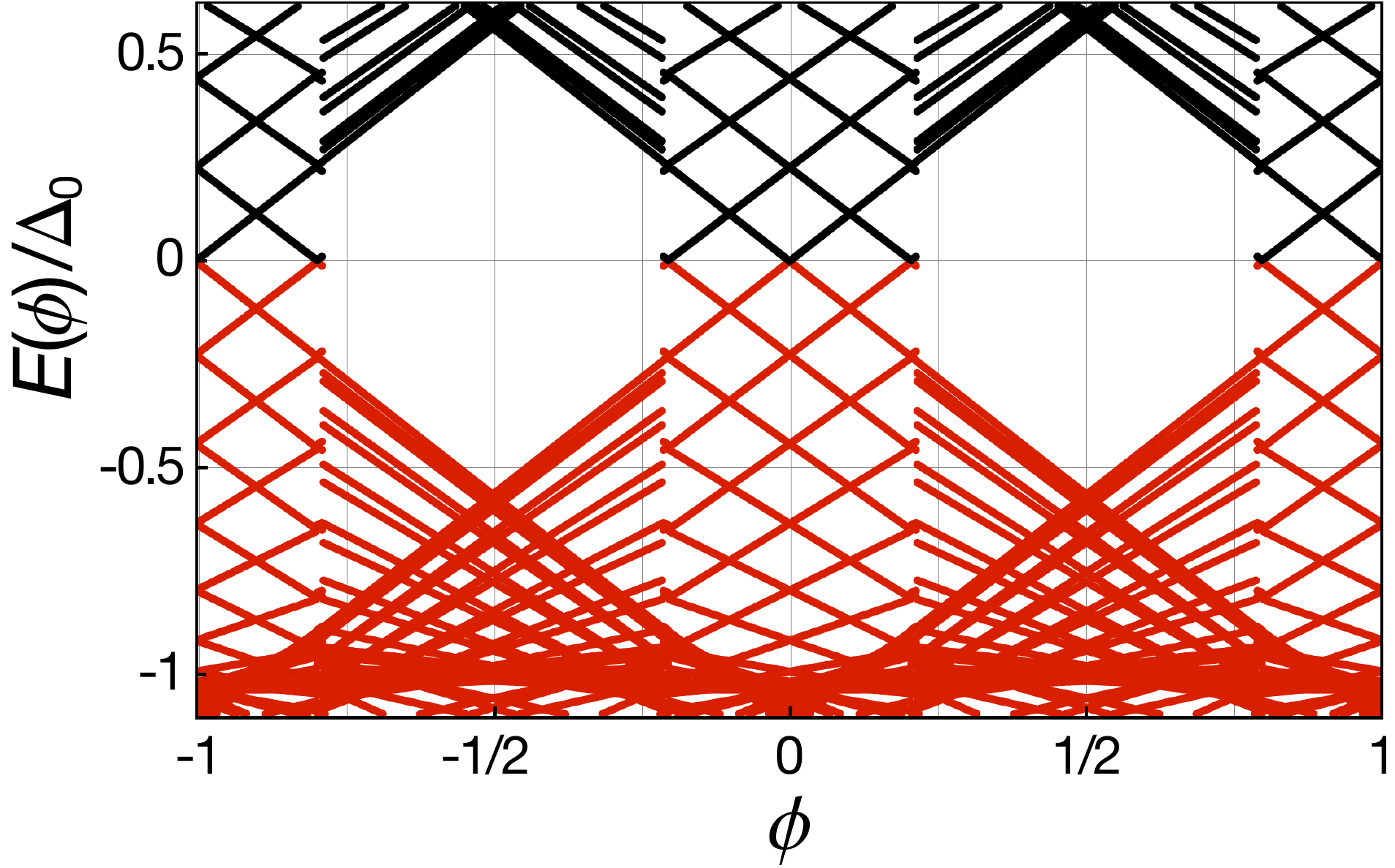}
\put(0,58){({\bf a})}
\end{overpic}
\begin{overpic}
[width=0.49\columnwidth]{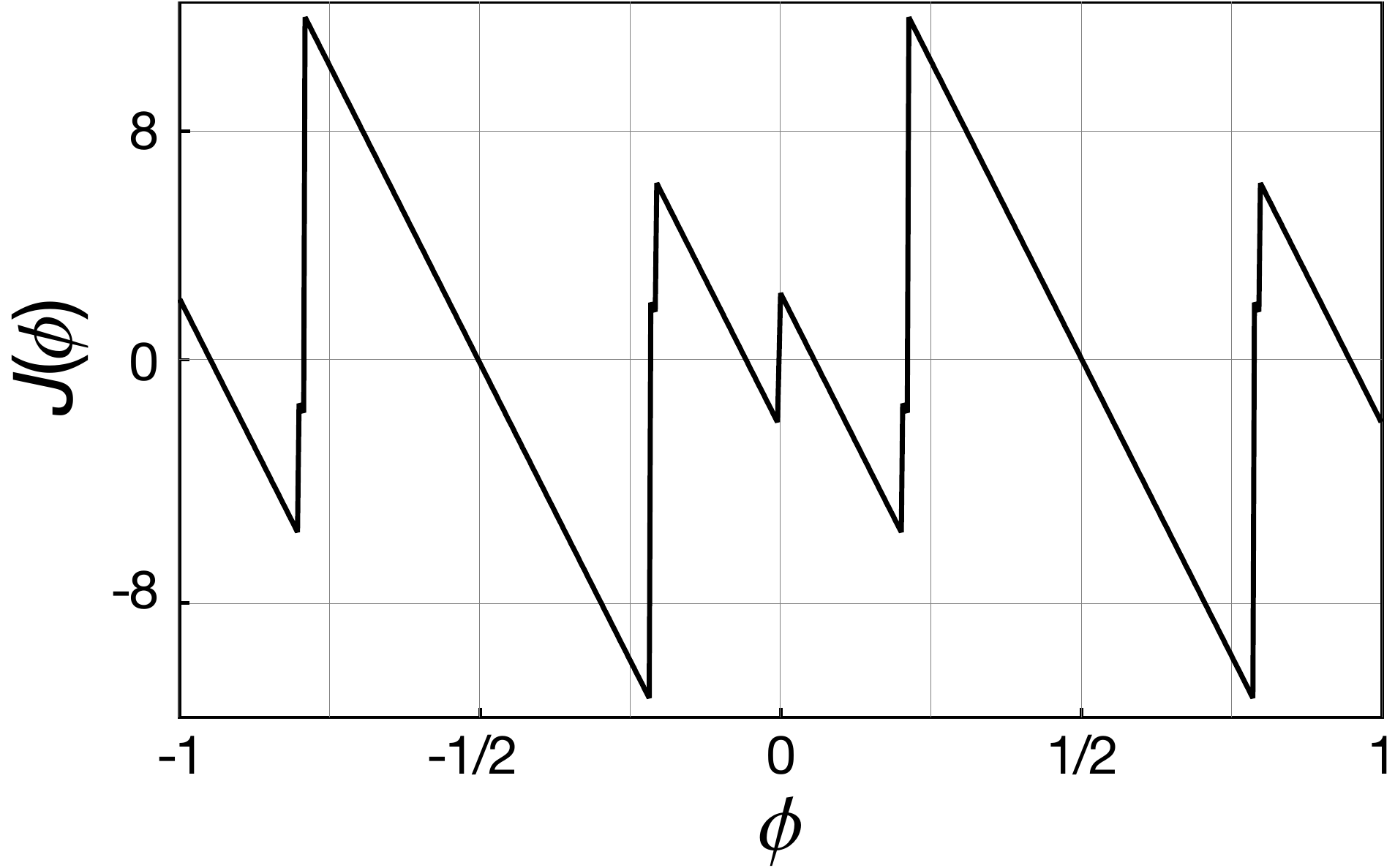}
\put(0,58){({\bf b})}
\end{overpic}
\caption{Energy spectrum and supercurrent $J_q(\phi)$ in a cylinder with circumference of $N=56$ and height $M=28$ and $\Delta_0=\Delta_q(\phi=0)\approx0.1t$. The red lines represent occupied states which form the condensate, whereas the black lines represent empty states. The spectrum (a) is similar to the one obtained in \cite{loder:08} for a square frame geometry. Clearly visible is the energy gap in the odd-$q$ flux sectors, whereas in the even-$q$ flux sectors states cross the Fermi energy upon changing the flux. At these crossing points, a jump in the supercurrent is observed (b).}
\label{Fig3}
\end{figure}

Loops of $d$-wave superconductors can be arranged in two different ways. In a first choice for the geometry the order parameter windes jointly with the lattice around a hole such that the phase of the order parameter remains constant on the selected path. The cylinder geometry described here is an example for this choice. The second option is to fix the orientation of the lattice and to cut out a hole. Then the phase of the order parameter rotates by $2\pi$ on any closed path encircling the hole once. This was investigated with a square frame in \cite{loder:08} and also in the 1D model in \cite{barash:07}. These two arrangements are in fact physically equivalent. The square frame geometry ensures the right number of lattice sites for the maximum difference in the spectrum of the even-$q$ and odd-$q$ flux sectors, as discussed in section~\ref{sec2}. For a direct comparison to the cylinder geometry, we chose a cylinder with $N=56$ and $M=28$, which has the same hole diameter as the square frame in \cite{loder:08}, and the ratio $N/M=2$ produces qualitatively the same energy spectrum. The resulting spectrum is shown in figure~\ref{Fig3}(a). It has indeed the same characteristic features as in the square frame geometry. An energy gap of the same order of magnitude exists in the odd-$q$ flux sectors, and the DOS in the even-$q$ flux sectors is gapless. There are no hybridization effects in the spectrum of the cylinder, since it preserves the full rotational symmetry. The features mentioned above are also in agreement with the qualitative discussion of section~\ref{sec3}. Clearly visible in figure~\ref{Fig3}(b) are the jumps in the supercurrent whenever an energy level crosses $E_F$, and the offset in the flux value for which $q$ changes (large jumps). This offset depends in a complex way on the system size and the pairing potential strength, but generally decreases for larger values of $N$, $M$, and $V$.

\begin{figure}[t]
\centering
\subfigure[$\ V=0.4t$]{\includegraphics[height=40mm]{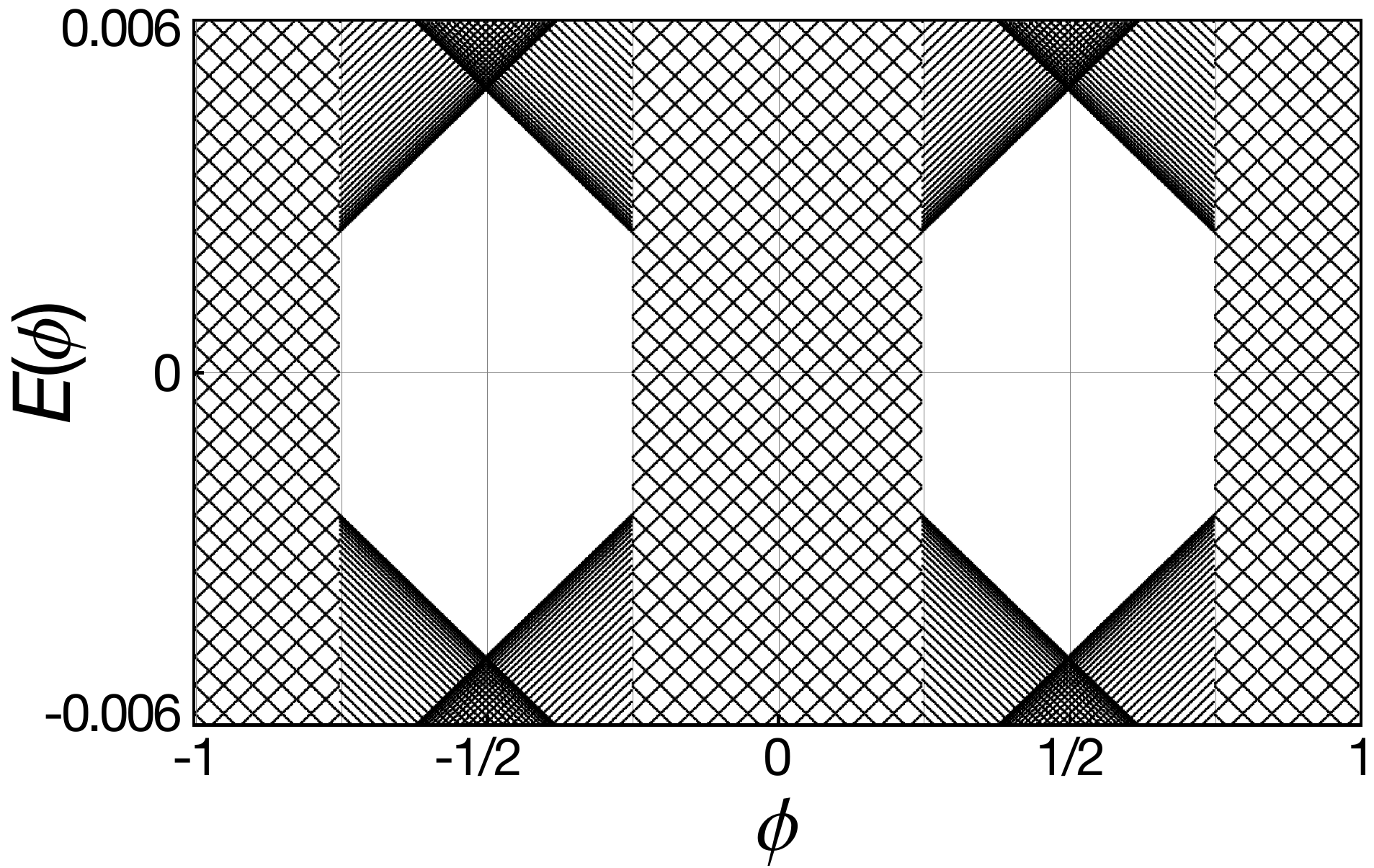}}
\subfigure[$\ V=0.2t$]{\includegraphics[height=40mm]{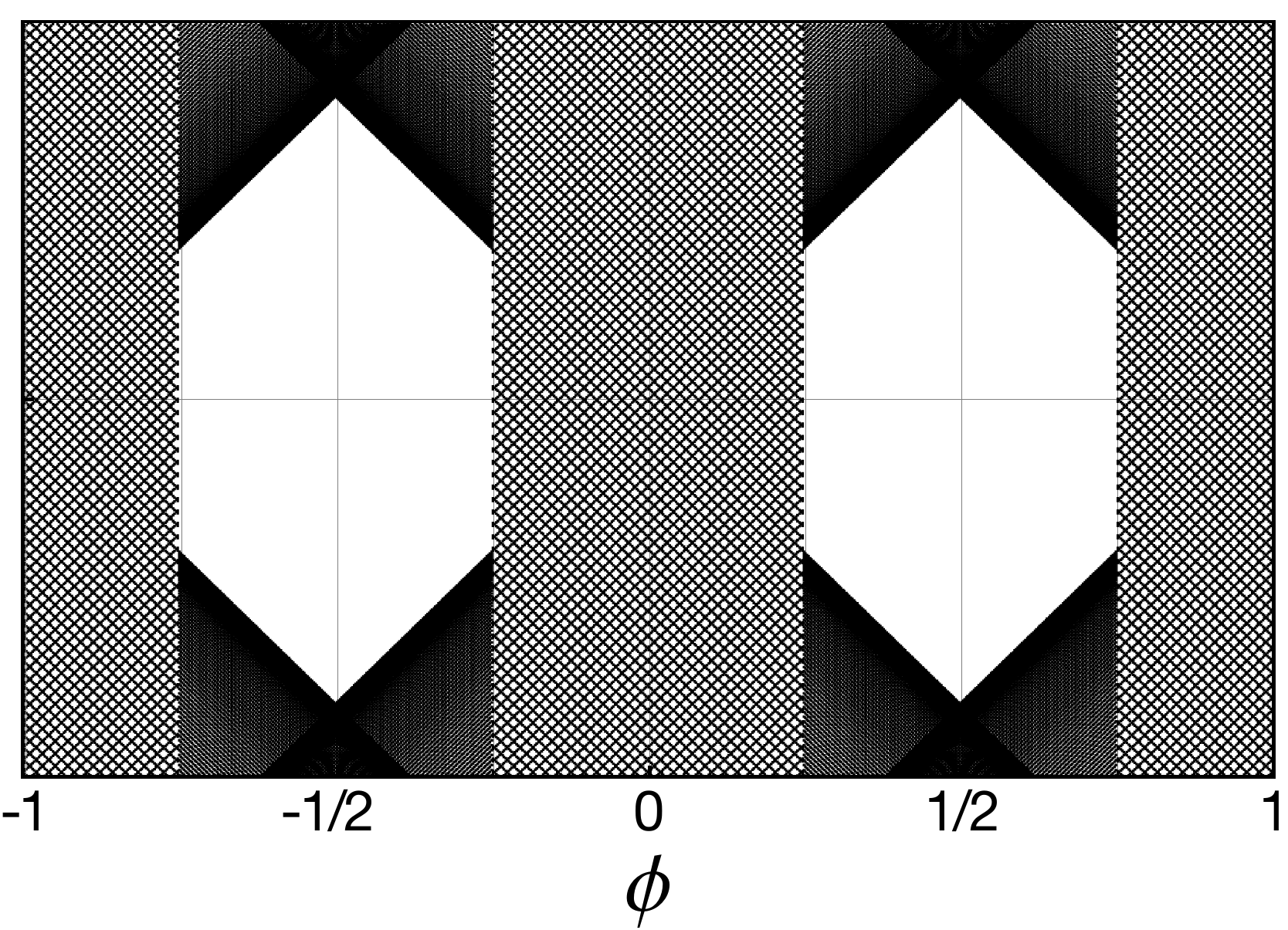}}
\subfigure[$\ V=0.4t$]{\includegraphics[height=40mm]{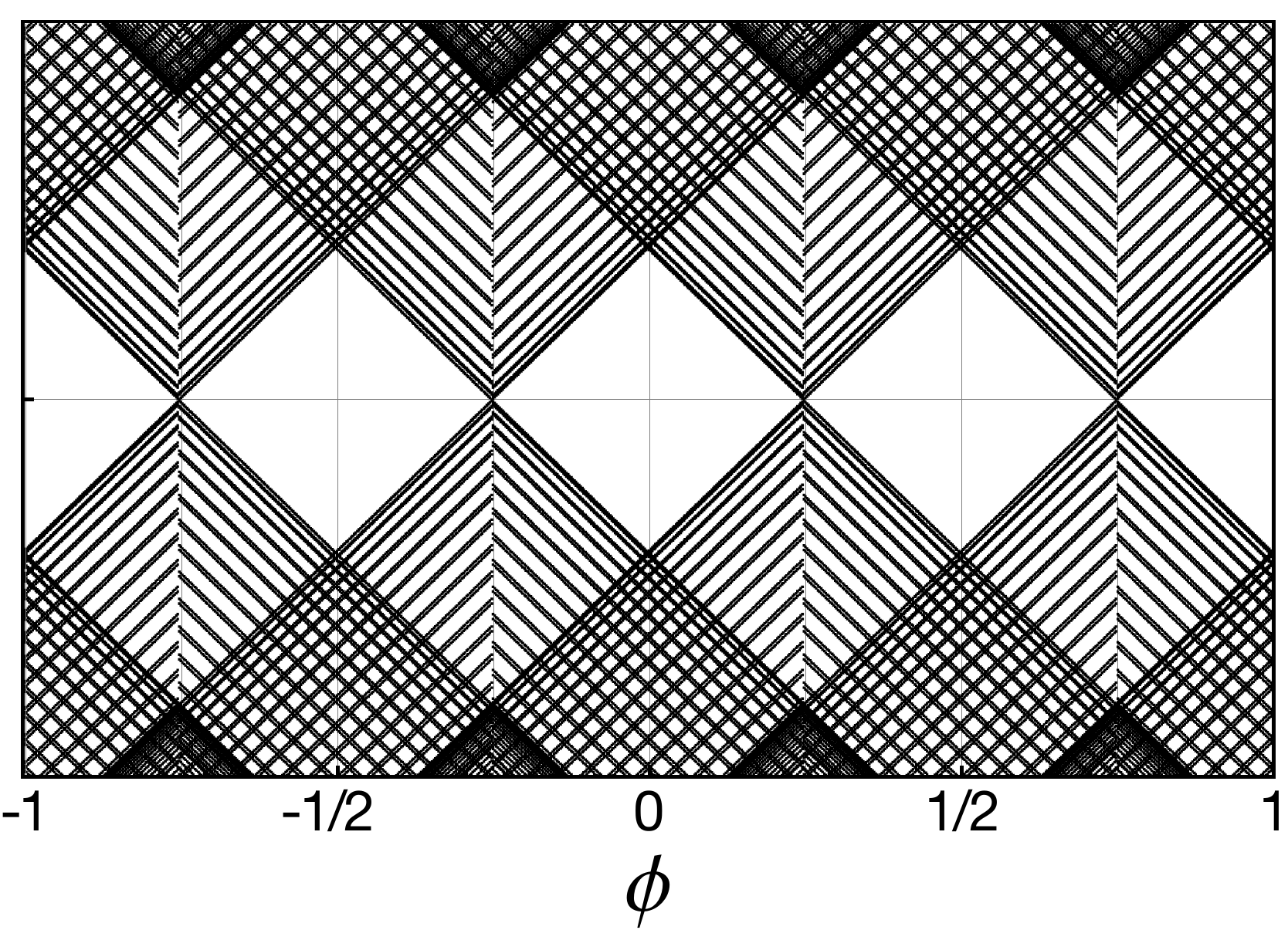}}
\caption{A section of the energy spectrum around $E_F$. (a) and (b) show the spectra of a cylinder with $N=M=2600$. (a) $V=0.4t$ and $\Delta_0(0)\approx0.05t$; (b) $V=0.2t$ and $\Delta_0(0)\approx0.02t$. The energy gap $l_1\ll\Delta_q(\phi)$ for these systems and all the states shown have the same Doppler shift (all lines are parallel / perpendicular). The density of states is quasi continuous in the even-$q$ flux sectors and grows linearly with decreasing $\Delta_q(\phi)$. (c) shows the $hc/2e$-periodic spectrum of a cylinder with $N=2600$ and $M=N+1$.}
\label{Fig4}
\end{figure}

The spectrum and the supercurrent in figure~\ref{Fig3} display the expected signatures of discreteness which are not captured by the analytic analysis of section~\ref{sec3}. The important parameter is obviously the size of the level spacing. Explicitly we take a closer look at a cylinder with $N=M=2600$, and thus a circumference of the order of 1 $\mu$m. The calculated spectra are shown in figure~\ref{Fig4}(a) and (b) for different pairing potentials, resulting in (a) $\Delta_q(\phi)\approx0.05t$ and (b) $\Delta_q(\phi)\approx0.02t$. The qualitative features ot the much smaller cylinder remain, but the gap $l_1$ in the odd-$q$ flux sectors is smaller, because $l_1$ decreases with $1/N$. In the even--$q$ flux sectors, there are $M$ levels spread out between $-\Delta_q(\phi)$ and $\Delta_q(\phi)$, which leads to an increase in the DOS around $E_F$ with decreasing $\Delta_q(\phi)$ for fixed $N$ and $M$. The representation with a continuous DOS is therefore appropriate for $\Delta_q(\phi)\ll t$, which is fulfilled well in figure~\ref{Fig4}(b).  Figure \ref{Fig4}(c) shows the spectrum for $M-1=N=2600$, which is almost identical in even-$q$ and odd-$q$ flux sectors. There is still a gap for non-integer (or half-integer) values of $\phi$, but it is equally distributed in the even-$q$ and odd-$q$ sectors. Other choices of $N$ and $M$ produce mixed features of the spectra in figure~\ref{Fig4}(a) and \ref{Fig4}(c). All the energy levels shown in each part of figure~\ref{Fig4}, which belong to nodal states,  have apparently the same derivative with respect to $\phi$.

\begin{figure}[t]
\centering
\begin{overpic}
[width=0.49\columnwidth]{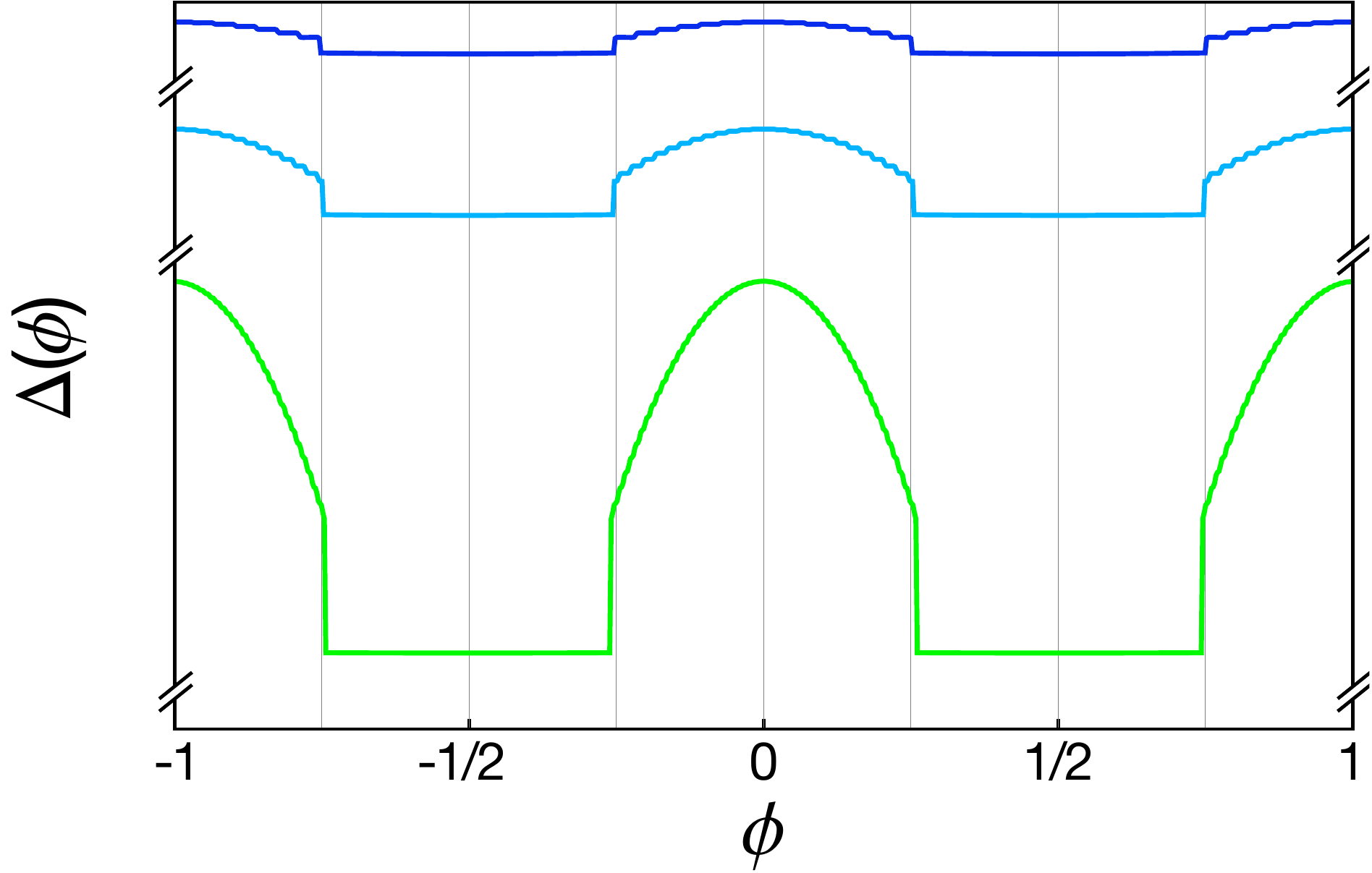}
\put(-1,60){({\bf a})}
\end{overpic}
\begin{overpic}
[width=0.49\columnwidth]{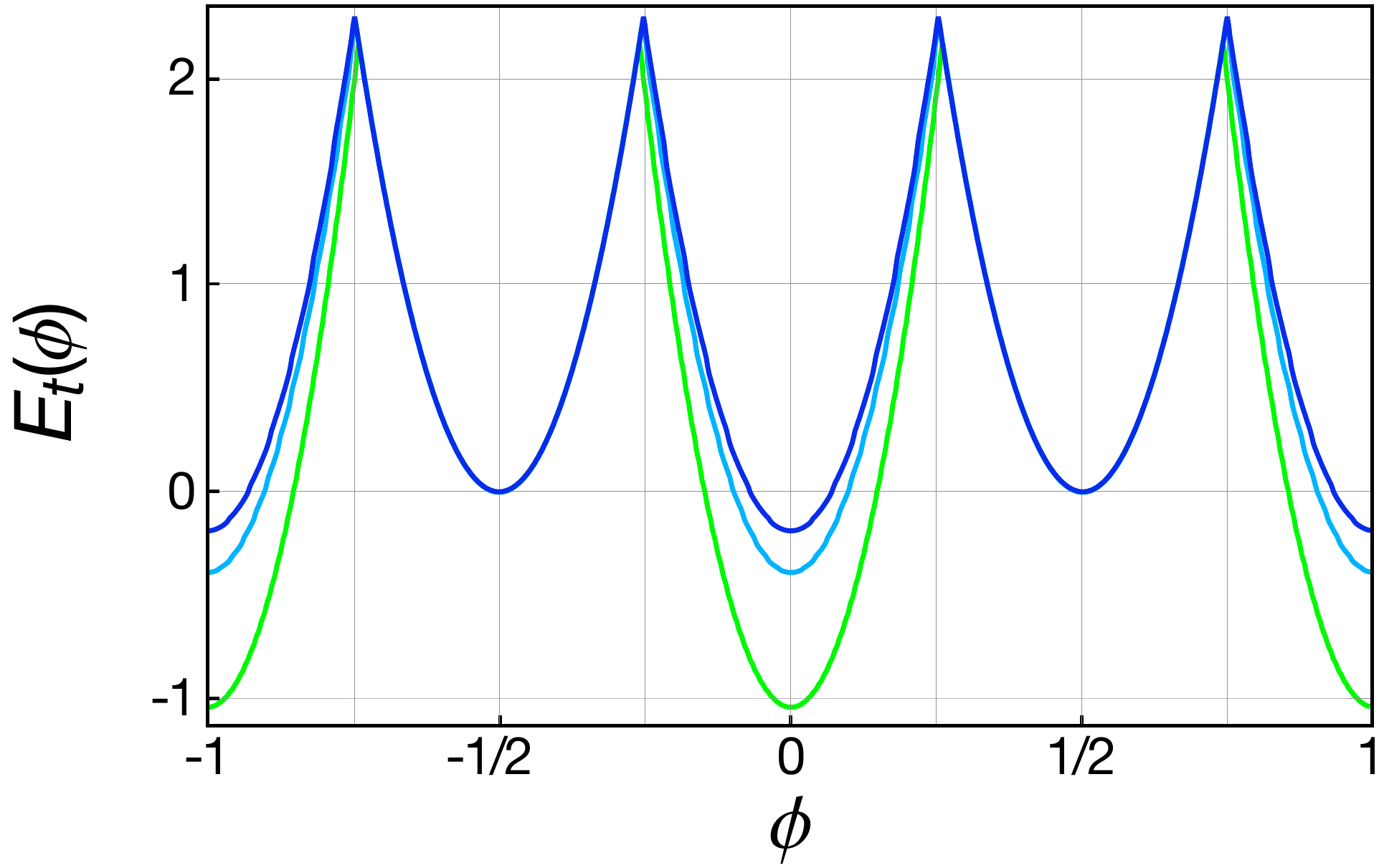}
\put(0,60){({\bf b})}
\end{overpic}\\[2mm]
\begin{overpic}
[width=0.49\columnwidth]{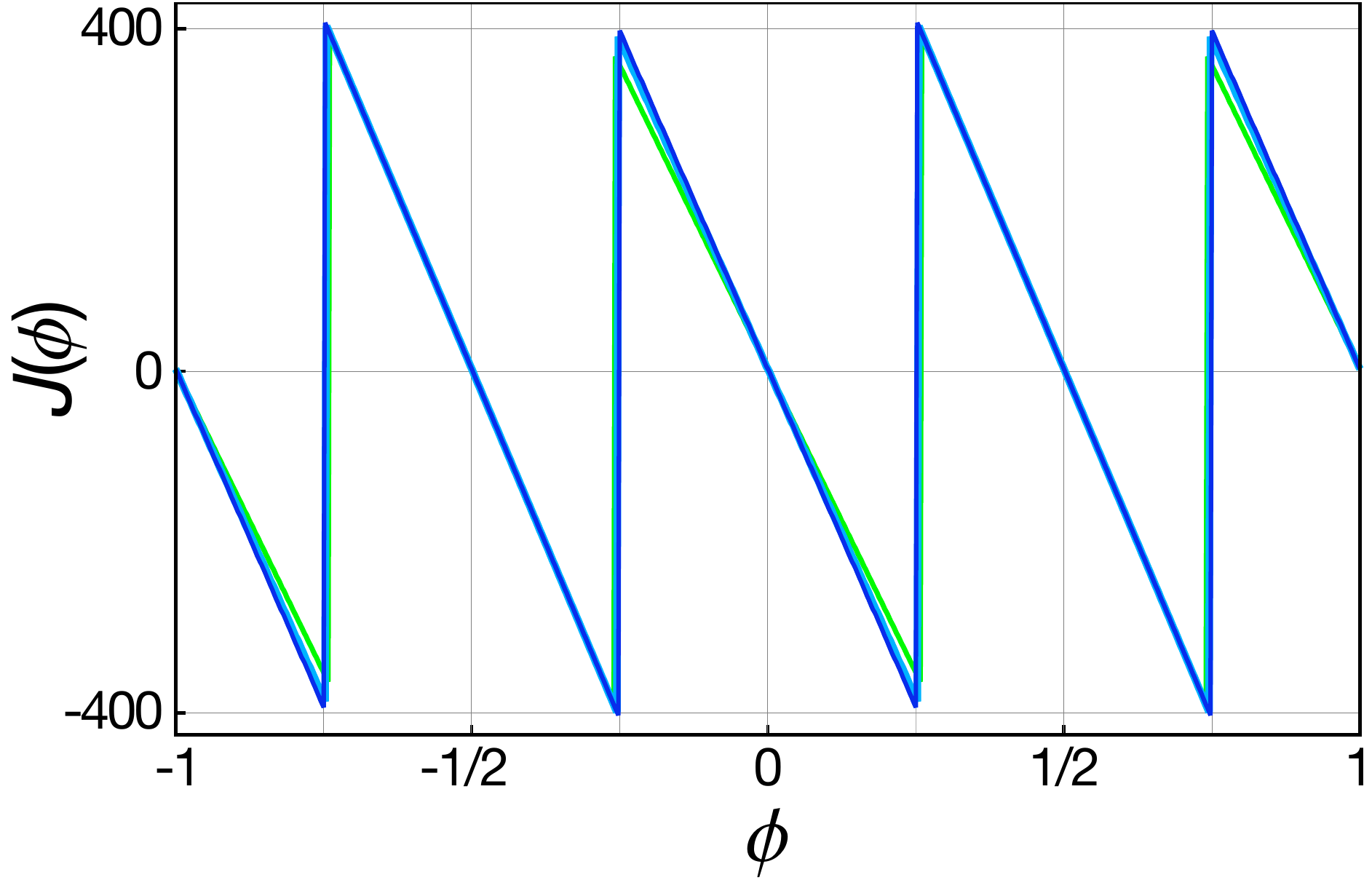}
\put(-1,60){({\bf c})}
\end{overpic}
\begin{overpic}
[width=0.49\columnwidth]{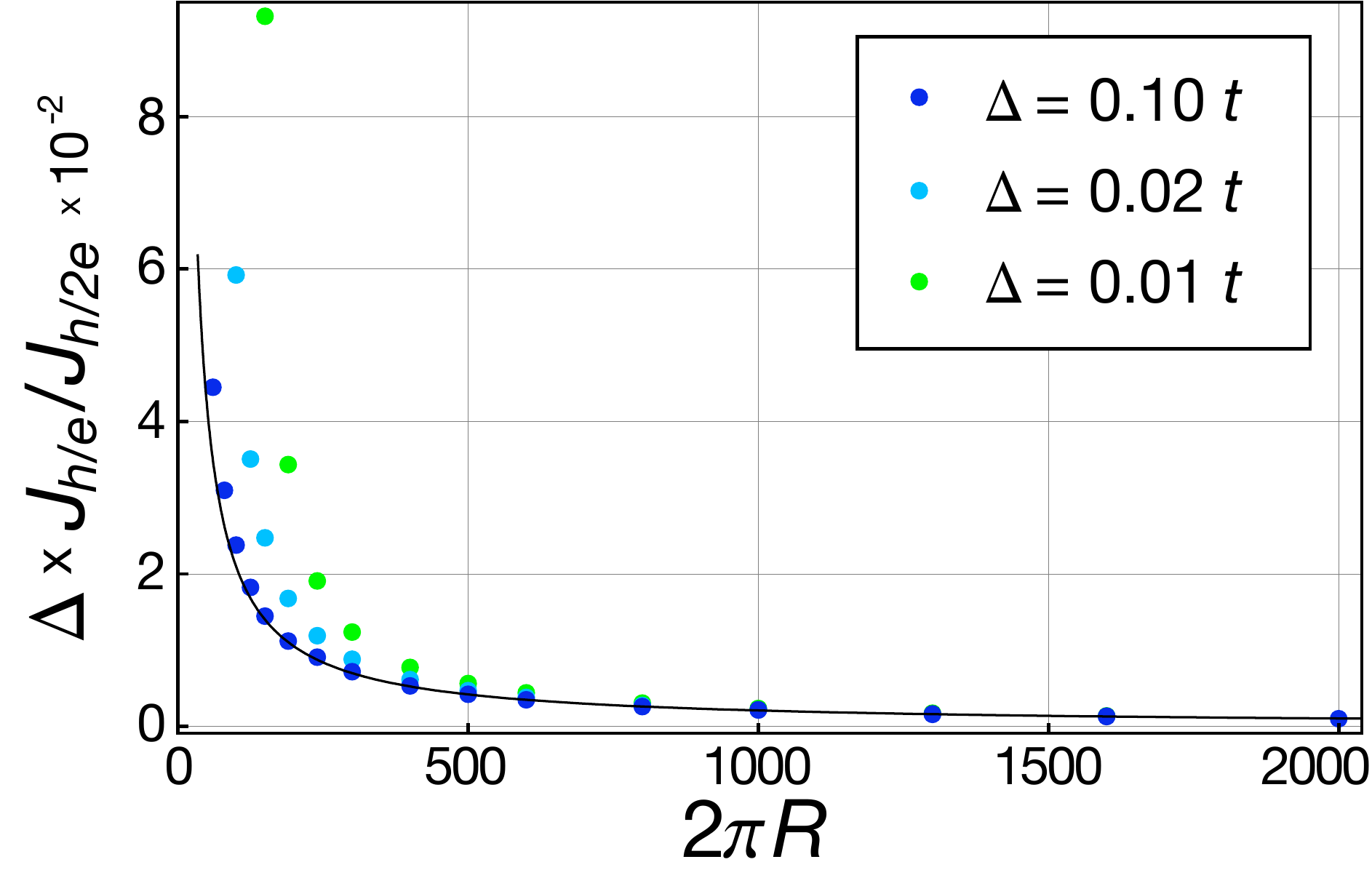}
\put(0,60){({\bf d})}
\end{overpic}
\caption{(a-c) Self-consistent order parameter $\Delta_q(\phi)$, total energy $E_t(\phi)$ and supercurrent $J_q(\phi)$ plotted as a function of flux $\phi$ for $V=0.25t$ (dark blue), $V=0.2t$ (light blue), and $V=0.15t$ (green). $\Delta_q(\phi)$ and $E_t(\phi)$ are shown in units of $t$ and $J_q(\phi)$ in units of $et/hc$. The oscillations of all quantities are $\propto1/R$ and of the order of $t$ for $E_t(\phi)$. The amplitude of the oscillations in $\Delta_q(\phi)$ are rather small. For $V=0.25t$, $\Delta_q(\phi)\approx0.036t$ with an oscillation amplitude $\delta\Delta=[\Delta_0(0)-\Delta_1(1/2)]/\Delta_0(0)\approx10^{-8}$, $\Delta_q(\phi)\approx0.02t$ and $\delta\Delta\approx5\times10^{-6}$ ($V=0.2t$), and $\Delta_q(\phi)\approx0.009t$ and $\delta\Delta\approx4\times10^{-5}$ ($V=0.15t$).
(d) Ratio of the first and second Fourier components of the supercurrent as a function of the cylinder radius $R$ for fixed values of $\Delta$. The  height $M$ of the cylinder is equal to $N=2\pi R$, which yields the maximum values for $J_{h/e}$. For $N$ larger than some $\Delta$-dependent number (see main text), the results of the exact evaluations fit very well to the prediction of equation~(\ref{s51}) (black line).}
\label{Fig5}
\end{figure}

The small level spacing in the $\mu$m sized cylinders results in solutions $\Delta_q(\phi)$ of the gap equation (\ref{s401}), which are nearly constant (figure~\ref{Fig5}(a), note the vertical scale discussed in the figure caption). The $\phi$-dependence of the total energy $E_t(q,\phi)$ also becomes small, whereas the small difference for even-$q$ and odd-$q$ remains important for the supercurrent $J_q(\phi)$. Since $J_q(\phi)\propto\partial E_t(q,\phi)/\partial\phi$, the differences in $E_t(q,\phi)$ imply different current amplitudes in the even and odd $q$ sectors [see figure~\ref{Fig5}(b) and (c)]. This effect is larger for smaller $\Delta_q(\phi)$, because the number of energy levels crossing $E_F$ increases with decreasing $\Delta_q(\phi)$. For the chosen pairing potentials $V$, the difference of the amplitudes of $J_q(\phi)$ for even and odd $q$ are of the order of a few percent. Per contra, the current jumps within a $q$-sector are tiny for the large radius of the $\mu$m-size cylinders. However, the resulting $\Delta_q(\phi)$ is considerably larger than in the $d$-wave cuprate superconductors. 
\footnote{Angle-resolved photoemission spectroscopy on various cuprates suggests a tight-binding $t\approx200$ meV -- 400 meV. The gap at the antinodes, obtained from tunneling spectra, varies between 10 meV and 50 meV \cite{shen, fischer}, therefore $\Delta\approx0.002t$ -- $0.01t$.}
Consequently, the upper limit for the difference of $J_q(\phi)$ in the even-$q$ and odd-$q$ sectors would be larger in the cuprate superconductors than in the model system calculated here. The offset in the jump of $J_q(\phi)$ from the flux values $\phi=(2n-1)/4$ is resolved for the smallest $V$ in figure~\ref{Fig5}. However, at low temperatures the superconducting state for each $q$ becomes meta stable for those flux values, for which it is not the ground state. At which flux values such a meta stable state decays into the ground state is not clear and the position of the jump in the supercurrent can vary in experiments.

We now compare the $R$-dependence of the ratio of the first and second Fourier components $J_{h/e}/J_{h/2e}$ analogous to section~\ref{sec3}. This is shown in figure~\ref{Fig5}(d) for different values of $V$. The ratio is in excellent agreement with equation~(\ref{s51}) for system sizes larger than a few hundred lattice constants. For smaller systems, $J_{h/e}$ becomes larger than predicted by the $1/R$ size dependence. The scale which decides about the validity of the approximations used in section~\ref{sec3} is the ratio of the level spacing and $\Delta_q(\phi)$. Equation~(\ref{s51}) therefore holds, if the prefactor of  $\phi$ in equation~(\ref{s45}) is small, that is, if $\sqrt{8}\,t\ll\pi\Delta R$, because $t/R$ is proportional to the level spacing of the nodal states. For a cylinder with radius $Ra=2600a\approx$ 1 $\mu$m and $\Delta_q(\phi)/t\approx0.01$, we obtain the ratio $J_{h/e}/J_{h/2e}\approx 0.04$, which is almost identical with the result of section~\ref{sec3}.

\section{Periodicity Crossover for small $\bm\Delta$} \label{sec5}
\begin{figure}[t]
\centering
\begin{overpic}
[width=0.49\columnwidth]{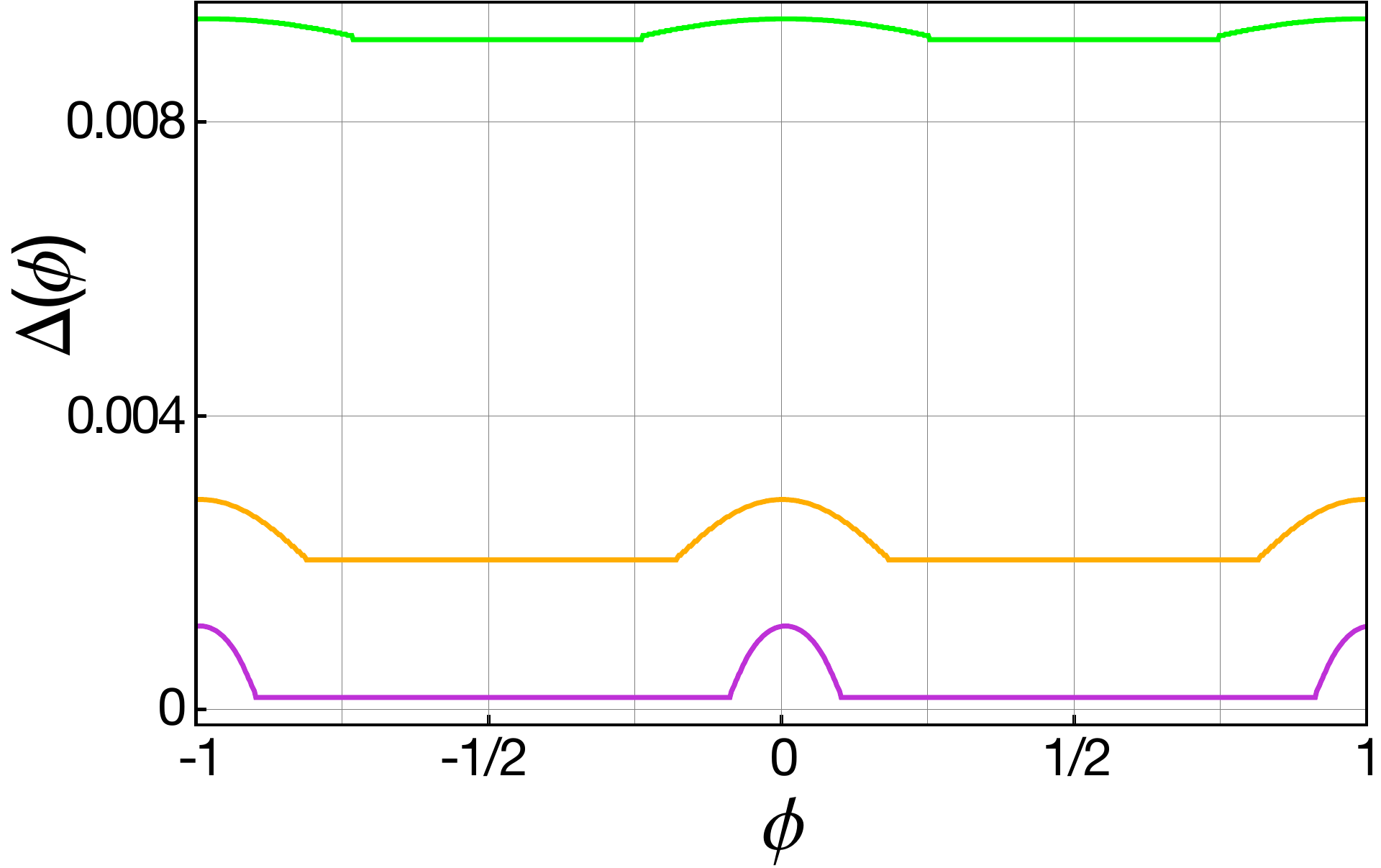}
\put(0,59.5){({\bf a})}
\end{overpic}
\begin{overpic}
[width=0.49\columnwidth]{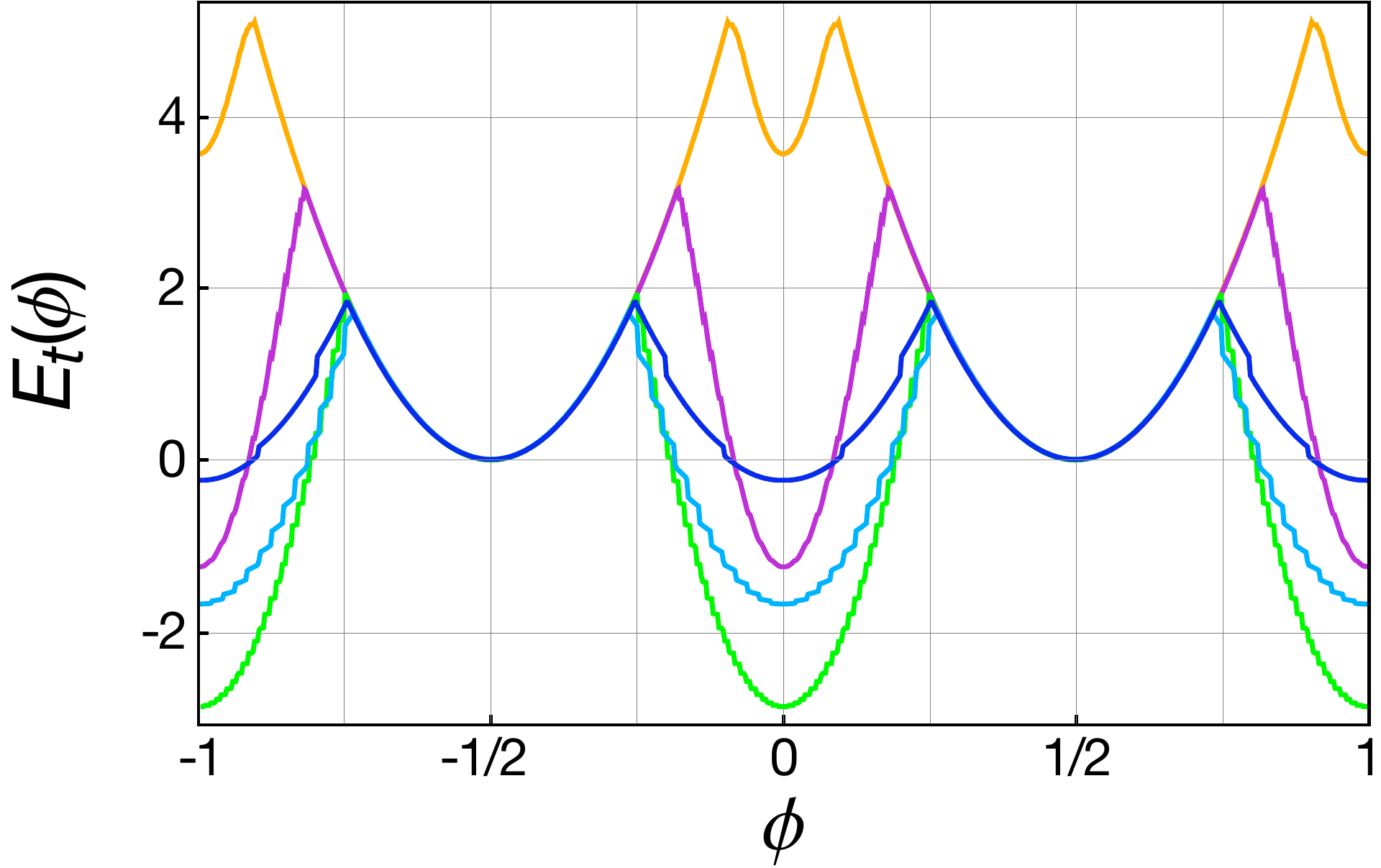}
\put(0,59.5){({\bf b})}
\end{overpic}\\[2mm]
\begin{overpic}
[width=0.49\columnwidth]{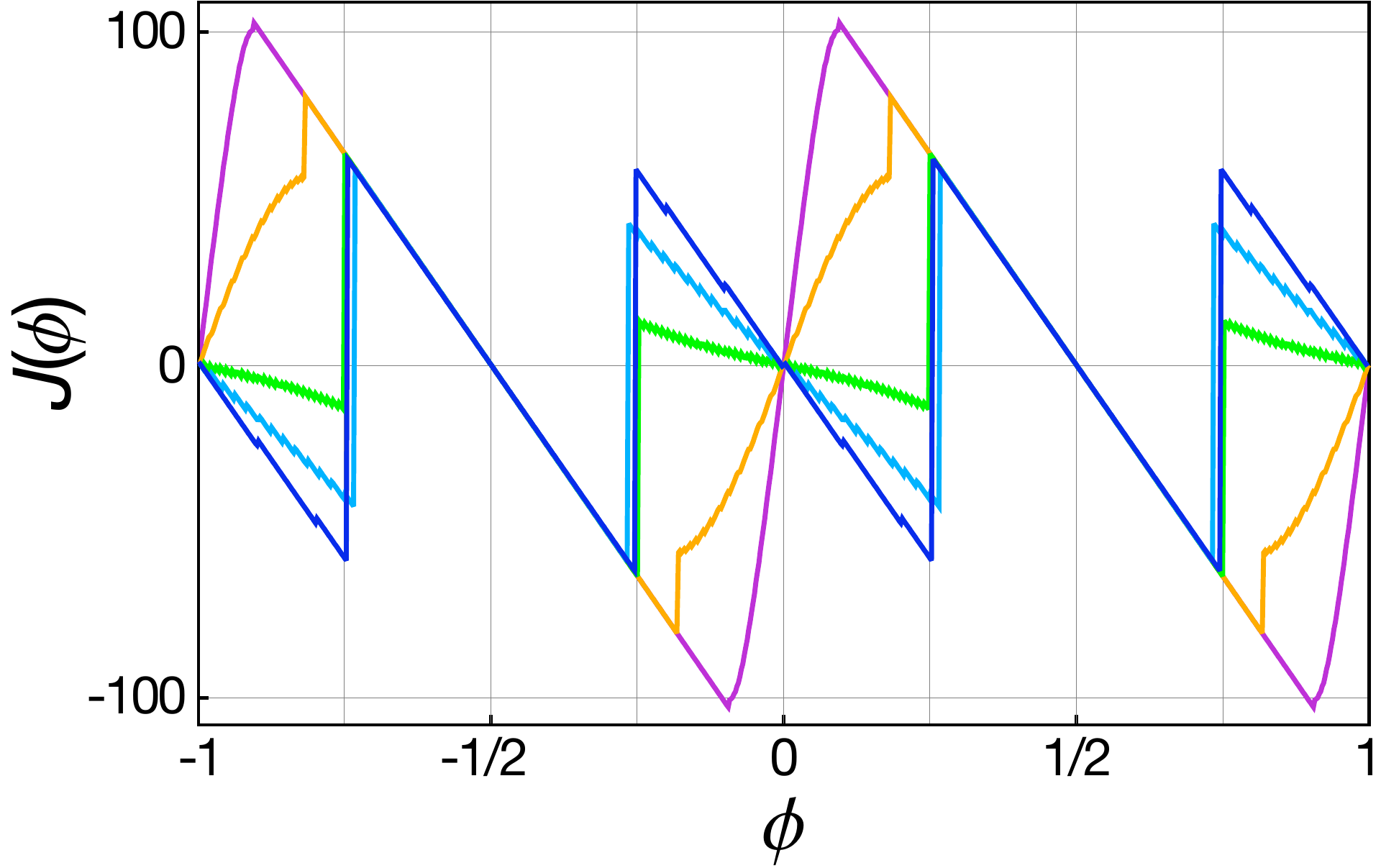}
\put(0,59.5){({\bf c})}
\end{overpic}
\caption{Periodicity crossover at $T=0$ in a cylinder with $N=M=400$ for $V=0.40t$ (dark blue), $V=0.2t$ (light blue), $V=0.15t$ (green), $V=0.1t$ (orange) and $V=0.07t$ (purple). For $V=0.1t$ and $V=0.07t$, $\Delta_q(\phi)$ is smaller than the Doppler shift for all $\phi$, thus $J_q(\phi)$ approaches the $hc/e$-periodic normal persistent current.}
\label{Fig7}
\end{figure}

So far we always assumed that $\Delta_q(\phi)\gg e_q(\phi)$ and concluded that variations in $\Delta_q(\phi)$ are negligible. But if $\Delta_q(\phi)$ is of the same order as the Doppler shift $e_q(\phi)$, the situation changes dramatically. This is the case, if either the radius $R$ of the cylinder is very small, or the pairing potential $V$ is small or the temperature $T$ is close to $T_c$. Here, we analyze the flux periodicity and the crossover from a \textquotedblleft small-gap\textquotedblright\ to a \textquotedblleft large gap\textquotedblleft\ regime by increasing $V$ from zero to higher values at $T=0$, and by lowering $T$ through $T_c$ for fixed $V$. As mentioned above, the amplitude of the oscillations, especially those of $\Delta_q(\phi)$, become very small for increasing $R$. For very large $R$, the periodicity crossover takes place  within a tiny range of $V$ or $T$, respectively. To observe the crossover more comfortably in a larger window of $V$ or $T$, we use smaller systems here.

The mechanism of the periodicity crossover at $T=0$, controlled by $V$, is best discussed by analyzing the total energy $E_t(q,\phi)$ [figure~\ref{Fig7}(b)]. It differs little from the crossover in $s$-wave superconductors, for which it was investigated in \cite{loder:08.2}. In the normal state ($V=0$), $E_t(q,\phi)$ is $q$-independent and consists of an $hc/e$ periodic series of parabolae. For increasing $V$, a new minimum in $E_t(q,\phi)$ forms at the crossing points of two parabolae. This minimum mover downward in energy until this new parabolic arc crosses the neighboring parabolae at the flux values $\phi=(2n-1)/4$. The energies of the old and the new minima are generally different for any finite system, but they approach each other when $\Delta_q(\phi)\gg\delta_F$. In the odd-$q$ flux sectors, $\Delta_q(\phi)$ is nearly constant because no energy levels cross $E_F$, whereas in the even-$q$ sectors, levels cross $E_F$ for all values of $V$. This causes the wiggles in $E_t(q,\phi)$ and the decrease of $\Delta_q(\phi)$ with increasing $\phi$ [figure~\ref{Fig7}(a)]. For the smallest two values shown in figure~\ref{Fig7}(a), $\Delta_q(\phi)$ approaches zero as a function of $\phi$ for even $q$; for this reason the odd-$q$ states extend far into the even-$q$ flux sectors. With increasing $V$, the nearly $hc/2e$ periodic sawtooth pattern of the supercurrent evolves from the $hc/e$ periodic normal persistent current [figure~\ref{Fig7}(c)].

\begin{figure}[t]
\centering
\begin{overpic}
[width=0.49\columnwidth]{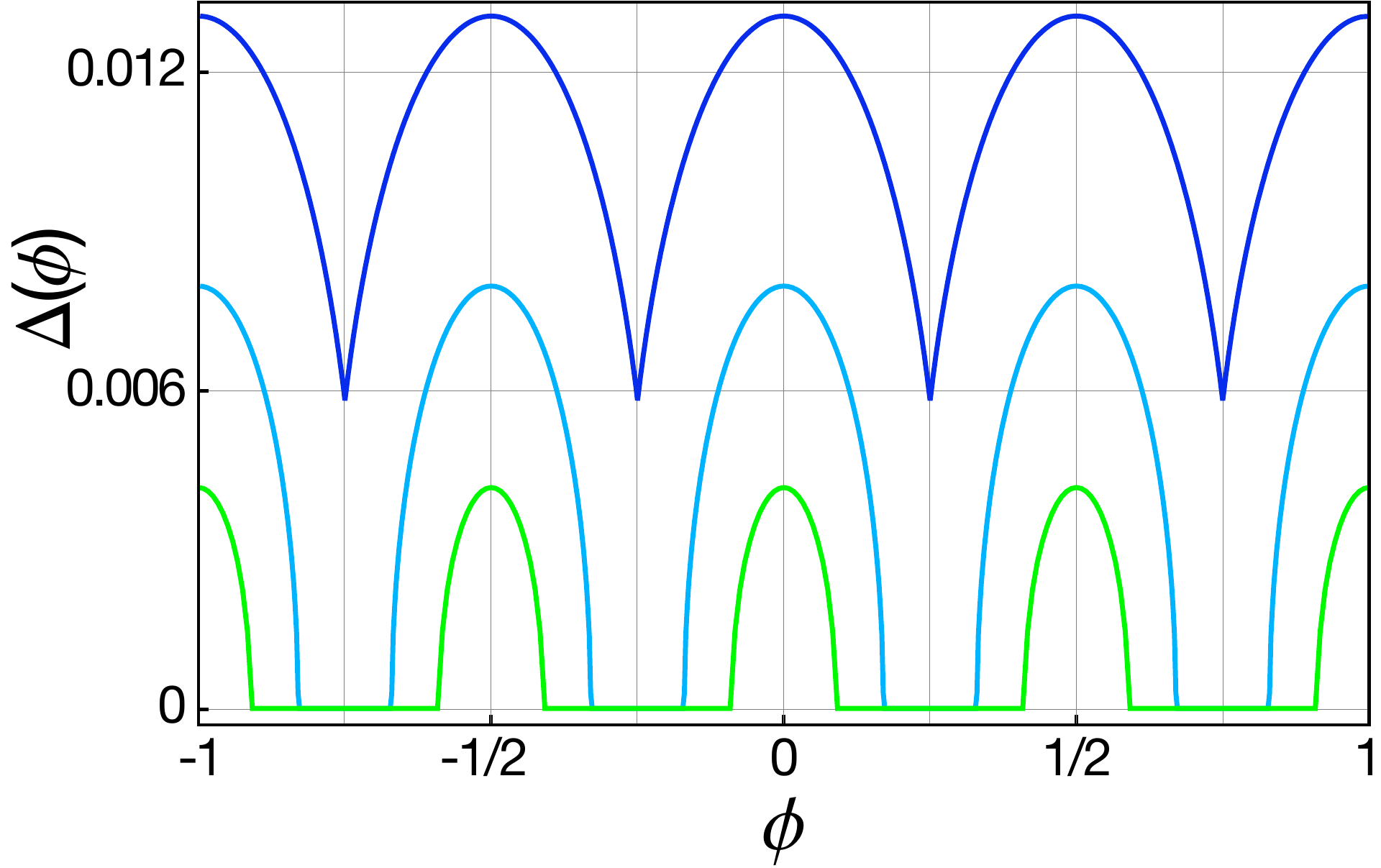}
\put(0,61.5){({\bf a})}
\end{overpic}
\begin{overpic}
[width=0.49\columnwidth]{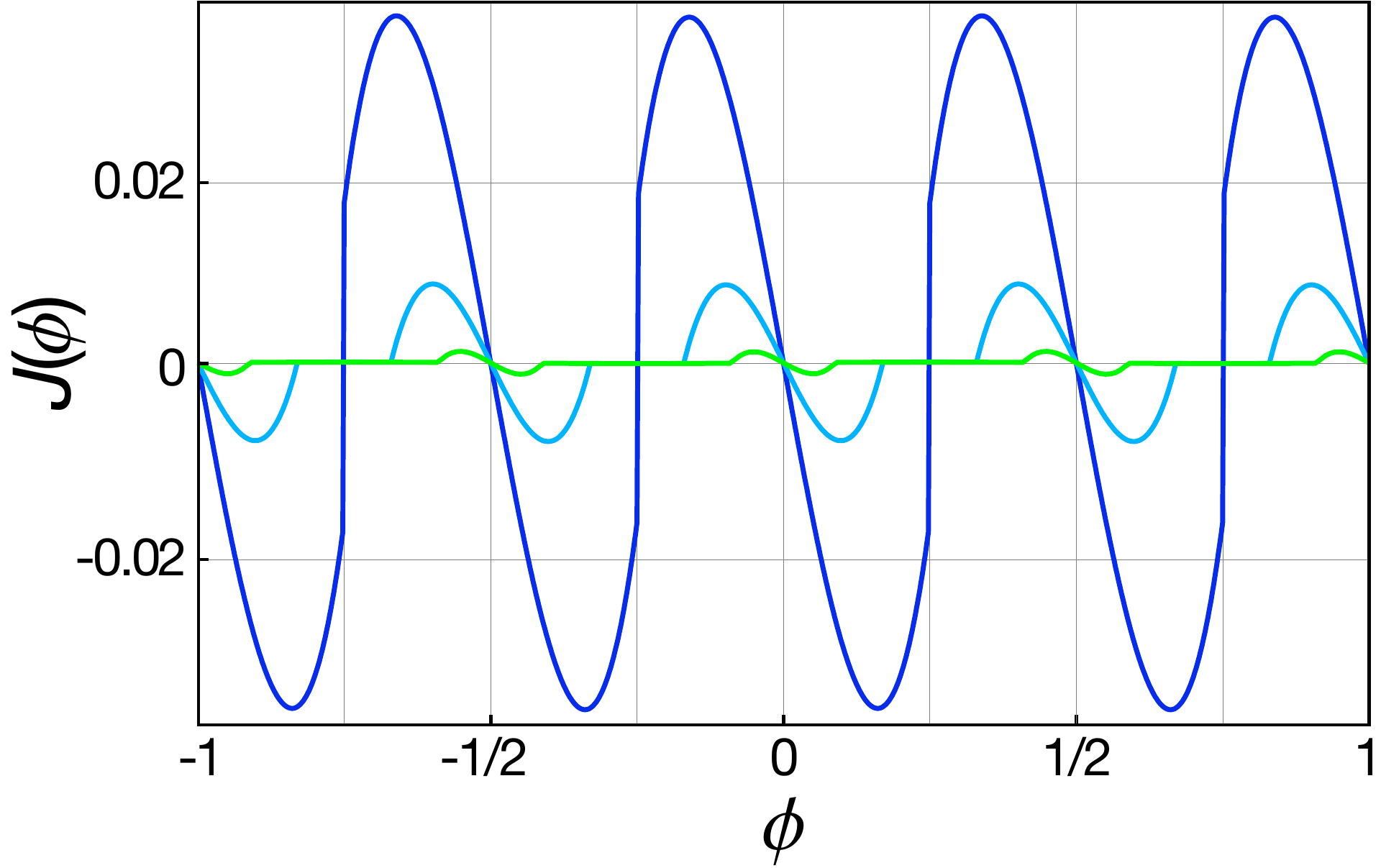}
\put(0,61.5){({\bf b})}
\end{overpic}
\caption{Temperature driven periodicity crossover for fixed $V=0.4t$ in a cylinder with $N=M=100$ for $k_{\rm B}T=0.1863t$ (dark blue), $k_{\rm B}T=0.1870t$ (light blue) and $k_{\rm B}T=0.1873t$ (green). The amplitude of the normal persistent current in the sectors with $\Delta_q(\phi)=0$ is much smaller than for $\Delta_q(\phi)>0$ and is invisible on this plot scale.}
\label{Fig8}
\end{figure}

The temperature controlled crossover at $T_c$ is analogous to the crossover controlled by $V$, but the finite temperature has quenched all the effects of discreteness as well as the gap in the odd-$q$ flux sectors. This means that the deviations from the $hc/2e$ periodicity are invisible in figure~\ref{Fig8}. Deviations appear with decreasing temperature as $k_{\rm B}T$ approaches $l_1$. The supercurrent decreases linearly with increasing $T$ until it reaches the exponentially small value of the normal persistent current at $T=T_c$ [figure~\ref{Fig8}(b)] \cite{vonoppen:92}. This suppression as well as the suppression of $\Delta_q(\phi)$ with temperature [figure~\ref{Fig8}(a)] differ only little from those of $s$-wave superconductors. The only qualitative difference is, that a characteristic temperature $T^*$ exists for $s$-wave superconductors, below which $\Delta(\phi=0)$ is larger than the maximum Doppler shift. This is equivalent to a coherence length $\xi(T^*)=2R$ \cite{loder:08.2}. Below $T^*$, $\Delta(\phi)>0$ for all $\phi$ in $s$-wave superconductors, and the thermodynamic quantities are therefore not affected by the Doppler shift. The relation $\Delta(\phi=0,T=0)\approx1.75T_c$ leads to the estimate
\begin{equation}
\frac{T_c-T^*}{T_c}\approx\frac{E_F}{3.1k_{\rm B}^2T_c^2R^2}.
\label{s403}
\end{equation}
For $d$-wave pairing, there is no such characteristic temperature because of the nodal states, but in analogy we can define $T^*$ as the crossover temperature below which $\Delta_q(\phi)>0$ for all $\phi$. Analogously to the $s$-wave case, we denote this situation as the \textquotedblleft large-gap\textquotedblright\ regime. For temperatures $T^*<T<T_c$, $\Delta_q(\phi)$ approaches zero for certain values of $\phi$, which we call the \textquotedblleft small-gap\textquotedblright\ regime. Since for a $d$-wave superconductor with nearest neighbor hopping $\Delta(\phi=0,T=0)>1.75k_{\rm B}T_c$ \cite{sigrist2}, one expects that $T_c-T^*$ is also larger and the crossover broader than for $s$-wave pairing.

\section{Conclusions} \label{sec6}
We have shown that in rings of unconventional superconductors with gap nodes, there is a paramagnetic, quasi-particle-like contribution $j_p>0$ to the supercurrent at $T=0$. This current is generated by the flux induced \textquotedblleft reoccupation\textquotedblright\ of nodal quasiparticle states slightly below and above $E_F$. Formally a coherence length $\hbar v_F/\Delta_\kv(\qv,\phi)>2R$ can be ascribed to these reoccupied states, which are therefore affected by the geometry of the system, however large the number of lattice sites is. If the normal state energy spectrum has a flux periodicity of $hc/e$, than the superconducting spectrum is $hc/e$ periodic, too. The normal state spectrum of a cylinder with a discrete lattice strongly depends on the number of lattice sites on the cylinder. This problem is characteristic for rotationally symmetric systems and is much less pronounced in geometries with lower symmetry, such as the square frame discussed in \cite{loder:08}. In such systems, the addition or removal of a  small number of lattice sites or impurities do not change the spectrum qualitatively, as tested by numerical calculations on a square frame. For an experimental arrangement where the difference in even and odd flux values is as large as possible, a square loop would be preferable.
Our results obtained in section~\ref{sec3} and section~\ref{sec4} for the periodicity of the physical quantities $\Delta_q(\phi)$, $E_t(q,\phi)$, and $J_q(\phi)$ provide therefore an upper limit for the $hc/e$ periodic components.

The $hc/e$ periodicity is best visible in the current component $j_p$ at $T=0$. For $d$-wave-pairing $j_p\propto1/R^2$, and the $hc/e$ periodic Fourier component decays like the inverse radius of the cylinder, relative to the $hc/2e$ periodic Fourier component. The lack of a characteristic length scale in nodal superconductors, such as  the coherence length for $s$-wave pairing, generates this algebraic decay with increasing $R$. Although $j_p$ is larger for small $\Delta$, it almost vanishes close to $T_c$, if $\Delta\gg\delta_F$, and variations of $T_c$ with flux, as in the Little-Parks experiment \cite{Little, Parks}, do not differ for $s$- and $d$-wave superconductors.

A possible set-up for the experimental detection of the $hc/e$ periodicity of the supercurrent is the insertion of Josephson junctions onto the cylinder, thereby creating a SQUID. The oscillations of the SQUID's critical current have the same flux periodicity as the circulating supercurrent. Indeed, experiments with $d$-wave SQUIDs by Schneider and Mannhart have shown an $hc/e$ periodic Fourier component under certain conditions \cite{schneider}. The relation to the effect described here however is not established, yet because of the so far unexplored influence of the Josephson junctions.

A different approach to study the cross over from the normal persistent current to the supercurrent in a ring was proposed by B\"uttiker and Klapwijk \cite{buttiker:86} and later by Cayssol \etal \cite{cayssol:03}. They analyzed a normal metal ring with an $s$-wave SC segment of variable length $l$. The energy spectrum, which they found, depends on $l$ in a similar way as it does in our analysis on the radius $R$. In this set-up, $hc/e$ periodicity should be found if $l<\xi_0$, although the ring diameter is much larger than $\xi_0$. Analogously, we expect the ratio $j_{h/e}/j_{h/2e}$ to be proportional to $1/l$ for a $d$-wave SC segment. This might be of advantage for experimental detection.

\ack

We are grateful to Yuri Barash and Doug Scalapino for helpful discussions in an early stage of this work and to Markus B\"uttiker for useful correspondence.
This work was supported by the Deutsche Forschungsgemeinschaft through SFB 484 and the EC (Nanoxide).

\end{document}